\input harvmac
\input epsf

\input labeldefs.tmp

\writedefs

\overfullrule=0pt

\def\Title#1#2{\rightline{#1}\ifx\answ\bigans
\nopagenumbers\pageno0\vskip1in
\else\pageno1\vskip.8in\fi \centerline{\titlefont #2}\vskip .5in}
\newcount\figno
\figno=0
\def\fig#1#2#3{
\par\begingroup\parindent=0pt\leftskip=1cm\rightskip=1cm\parindent=0pt
\baselineskip=11pt
\global\advance\figno by 1
\midinsert
\epsfxsize=#3
\centerline{\epsfbox{#2}} 
\vskip 12pt 
{\bf Figure\ \the\figno: } #1\par
\endinsert\endgroup\par
}
\def\figlabel#1{\xdef#1{\the\figno}}
\def\encadremath#1{\vbox{\hrule\hbox{\vrule\kern8pt\vbox{\kern8pt
\hbox{$\displaystyle #1$}\kern8pt}
\kern8pt\vrule}\hrule}}

\lref\HeemskerkPN{
  I.~Heemskerk, J.~Penedones, J.~Polchinski and J.~Sully,
  ``Holography from Conformal Field Theory,''
  JHEP {\bf 0910}, 079 (2009)
  [arXiv:0907.0151 [hep-th]].
}
\lref\HeemskerkTY{
  I.~Heemskerk and J.~Sully,
  ``More Holography from Conformal Field Theory,''
  JHEP {\bf 1009}, 099 (2010)
  [arXiv:1006.0976 [hep-th]].
}

\lref\FitzpatrickZM{
  A.~L.~Fitzpatrick, E.~Katz, D.~Poland and D.~Simmons-Duffin,
  ``Effective Conformal Theory and the Flat-Space Limit of AdS,''
  arXiv:1007.2412 [hep-th].
}

\lref\WeinbergCP{
  S.~Weinberg,
  ``The Cosmological Constant Problem,''
  Rev.\ Mod.\ Phys.\  {\bf 61}, 1 (1989).
}

\lref\PolchinskiGY{
  J.~Polchinski,
  ``The Cosmological Constant and the String Landscape,''
  arXiv:hep-th/0603249.
}

\lref\AharonySX{
  O.~Aharony, J.~Marsano, S.~Minwalla, K.~Papadodimas and M.~Van Raamsdonk,
  ``The Hagedorn / deconfinement phase transition in weakly coupled large N
  gauge theories,''
  Adv.\ Theor.\ Math.\ Phys.\  {\bf 8}, 603 (2004)
  [arXiv:hep-th/0310285].
}

\lref\FitzpatrickHH{
  A.~L.~Fitzpatrick and D.~Shih,
  ``Anomalous Dimensions of Non-Chiral Operators from AdS/CFT,''
  arXiv:1104.5013 [hep-th].
}

\lref\FanWM{
  J.~Fan,
  ``Effective AdS/renormalized CFT,''
  arXiv:1105.0678 [hep-th].
}

\lref\KachruYS{
  S.~Kachru and E.~Silverstein,
  ``4-D conformal theories and strings on orbifolds,''
  Phys.\ Rev.\ Lett.\  {\bf 80}, 4855 (1998)
  [arXiv:hep-th/9802183].
}

\lref\PolyakovAF{
  A.~M.~Polyakov,
  ``Gauge fields and space-time,''
  Int.\ J.\ Mod.\ Phys.\  A {\bf 17S1}, 119 (2002)
  [arXiv:hep-th/0110196].
}

\lref\SundborgUE{
  B.~Sundborg,
  ``The Hagedorn Transition, Deconfinement and N=4 SYM Theory,''
  Nucl.\ Phys.\  B {\bf 573}, 349 (2000)
  [arXiv:hep-th/9908001].
}

\lref\PenedonesUE{
  J.~Penedones,
  ``Writing CFT correlation functions as AdS scattering amplitudes,''
  JHEP {\bf 1103}, 025 (2011)
  [arXiv:1011.1485 [hep-th]].
}
\lref\KiritsisXC{
  E.~Kiritsis and V.~Niarchos,
  ``Large-N limits of 2d CFTs, Quivers and AdS3 duals,''
  JHEP {\bf 1104}, 113 (2011)
  [arXiv:1011.5900 [hep-th]].
}

\lref\ElShowkAG{
  S.~El-Showk and K.~Papadodimas,
  ``Emergent Spacetime and Holographic CFTs,''
  arXiv:1101.4163 [hep-th].
}
\lref\LiuTH{
  H.~Liu,
  ``Scattering in anti-de Sitter space and operator product expansion,''
  Phys.\ Rev.\  D {\bf 60}, 106005 (1999)
  [arXiv:hep-th/9811152].
}

\lref\GaberdielPZ{
  M.~R.~Gaberdiel and R.~Gopakumar,
  ``An AdS3 Dual for Minimal Model CFTs,''
  Phys.\ Rev.\  D {\bf 83}, 066007 (2011)
  [arXiv:1011.2986 [hep-th]].
}

\lref\RattazziPE{
  R.~Rattazzi, V.~S.~Rychkov, E.~Tonni and A.~Vichi,
  ``Bounding scalar operator dimensions in 4D CFT,''
  JHEP {\bf 0812}, 031 (2008)
  [arXiv:0807.0004 [hep-th]].
}

\lref\DolanUT{
  F.~A.~Dolan and H.~Osborn,
  ``Conformal four point functions and the operator product expansion,''
  Nucl.\ Phys.\  B {\bf 599}, 459 (2001)
  [arXiv:hep-th/0011040].
}

\lref\ElShowkAG{
  S.~El-Showk and K.~Papadodimas,
  ``Emergent Spacetime and Holographic CFTs,''
  arXiv:1101.4163 [hep-th].
}

\lref\KlebanovJA{
  I.~R.~Klebanov and A.~M.~Polyakov,
  ``AdS dual of the critical O(N) vector model,''
  Phys.\ Lett.\  B {\bf 550}, 213 (2002)
  [arXiv:hep-th/0210114].
}

\lref\WittenKH{
  E.~Witten,
  ``Baryons in the 1/n Expansion,''
  Nucl.\ Phys.\  B {\bf 160}, 57 (1979).
}

\lref\VolovikBH{
  G.~E.~Volovik,
  ``Vacuum Energy: Myths and Reality,''
  Int.\ J.\ Mod.\ Phys.\  D {\bf 15}, 1987 (2006)
  [arXiv:gr-qc/0604062].
}

\lref\GubserBC{
  S.~S.~Gubser, I.~R.~Klebanov and A.~M.~Polyakov,
  ``Gauge theory correlators from noncritical string theory,''
  Phys.\ Lett.\  B {\bf 428}, 105 (1998)
  [arXiv:hep-th/9802109].
}

\lref\WittenQJ{
  E.~Witten,
  ``Anti-de Sitter space and holography,''
  Adv.\ Theor.\ Math.\ Phys.\  {\bf 2}, 253 (1998)
  [arXiv:hep-th/9802150].
}

\lref\VerlindeHP{
  E.~P.~Verlinde,
  ``On the Origin of Gravity and the Laws of Newton,''
  JHEP {\bf 1104}, 029 (2011)
  [arXiv:1001.0785 [hep-th]].
}

\lref\MaldacenaRE{
  J.~M.~Maldacena,
  `The Large N limit of superconformal field theories and supergravity,''
  Adv.\ Theor.\ Math.\ Phys.\  {\bf 2}, 231 (1998)
  [Int.\ J.\ Theor.\ Phys.\  {\bf 38}, 1113 (1999)]
  [arXiv:hep-th/9711200].
}

\Title{\vbox{\baselineskip12pt
\hbox{arXiv:xxxx.xxxx}
\hbox{CERN-PH-TH/2011-}
}}
{\vbox{
\centerline{AdS/CFT and the cosmological constant problem}
\bigskip
}}
\centerline{Kyriakos Papadodimas}

\bigskip

 \centerline{Theory Group, Physics Department, CERN}
 \centerline{CH-1211 Geneva 23, Switzerland}
 \centerline{\it kyriakos.papadodimas@cern.ch}

\bigskip
\bigskip
\bigskip

\noindent  Within the context of the AdS/CFT correspondence we attempt to 
formulate the cosmological constant problem in
the dual conformal field theory. The fine-tuning of the bulk
cosmological constant is related to an apparent fine-tuning in the
effective description of the CFT in terms of its light operators:
while the correlators of single particle operators satisfy a large $N$
expansion, the expansion does not appear to be natural. Individual
terms contributing to correlators have parametrically larger value
than the one dictated by large $N$ counting.  The final $1/N$
suppression of correlators is achieved via delicate cancellations
between such terms. We speculate on the existence of underlying
principles which might make the bulk theory (secretly) natural.

\smallskip
\noindent
\Date{June , 2011}

\newsec{Introduction}

Does the holographic correspondence teach us anything new about the
cosmological constant problem? One of the most striking aspects of
holography is that spacetime and the quantum fields that we use to
describe it are emergent. This suggests the possibility that the
symmetry, or mechanism, responsible for the apparent fine-tuning of
the vacuum energy (and perhaps other fine-tuned couplings) may be
manifest in the underlying fundamental theory, but invisible, or
difficult to understand, in terms of the emergent light fields which
we directly observe.
\medskip
The AdS/CFT correspondence \MaldacenaRE,\GubserBC,\WittenQJ\ offers a
framework where this possibility can be investigated in a controlled
way.  In the real world we observe a positive cosmological constant
and (one of) the puzzle(s) is why the observed value is so small
compared to the cutoff scale of effective field
theory \WeinbergCP,\PolchinskiGY. However the puzzle is insensitive to
the {\it sign} of the cosmological constant. It would be equally
puzzling if the observed c.c. was of the same magnitude but opposite
sign. Hence it might be useful to try to understand the analogue of
the c.c. problem in anti-de Sitter space and within the context of
AdS/CFT. The latter provides us with a non-perturbative definition of
AdS quantum gravity in terms of a non-gravitational quantum field
theory, in which no mysteries about the vacuum energy and its
renormalization should exist.
\medskip
Our goal is to take a first step towards these questions by
considering the analogue of the cosmological constant problem in
anti-de Sitter space and translating it into some statement in the
dual CFT. Ultimately we would like to ask whether the CFT can resolve
the problem in a natural way.  While we have not been able to fully
reach these goals, we believe that this line of investigation may
eventually offer some new insights.
\medskip
More specifically in this paper we will try to explore the following
three questions:
\medskip
1. What are the properties of a CFT whose holographic dual exhibits a
``cosmological constant problem''?
\medskip
2. In such a theory how does the c.c. problem manifest itself in the
CFT? i.e.  does the CFT appear to be fine-tuned in some sense?\medskip
3. Does the CFT offer a resolution of the problem?
\bigskip
Let us give a brief summary of our conclusions: CFTs with holographic
duals are characterized by some sort of large $N$ expansion, which
means that correlators of single-particle operators are suppressed by
powers of $1/N$. Our main observation is that in CFTs whose bulk dual
exhibits a cosmological constant problem\foot{Assuming that such CFTs
exist.}, the large $N$ expansion does not appear to be natural, at
least not if we try to understand it in terms of the light
operators\foot{In this paper by ``light'' operators we mean those of
low conformal dimension.} of the CFT (in terms of glueballs and mesons, in
gauge theory language).  In this description it is reasonable to
expand the correlators into sums of terms, each of which corresponds
to the exchange of certain gauge invariant operators in intermediate
channels i.e. into a sum of  conformal blocks. While the sum of all
these contributions is suppressed by the expected power of $1/N$,
individual terms in the sum can be parametrically larger, as
long as they cancel among themselves. We believe that these surprising
cancellations  are the way that the bulk c.c. fine-tuning becomes
visible in the dual CFT.
\medskip
We also explore the idea that the fine-tuning may be resolved and that
the $1/N$ expansion may recover a natural form, once the correlators
are expressed directly in terms of the underlying fundamental fields
i.e. the ``quarks and gluons''. Analyzing the correlators in the
language of the gauge singlets described in the previous paragraph is
useful for certain purposes, for example in order to understand how
the bulk dual emerges, but may obscure other
properties of the theory. In particular, the $1/N$ suppression of
correlators may look natural in one description (in terms of
fundamental fields) but fine-tuned in the other (in terms of gauge
singlets). If this is true it means that the fine-tuning, both in the
bulk and the boundary, is an artifact of trying to understand the
correlation functions in terms of effective, rather than fundamental,
variables.
\medskip
\noindent {\it Some caveats}
\medskip
As we will explain later, we do not know at the moment a specific
large $N$ CFT whose dual exhibits a sharp cosmological constant
problem. Whether one exists or not is an interesting question,
equivalent to whether semi-classical AdS gravity without low energy
supersymmetry can be UV-completed. Nevertheless we will assume that
such a CFT exists and we will try to analyze what would be the CFT
manifestation of the bulk c.c. fine-tuning. Lacking a concrete example
to work with, our discussion will have to be somewhat abstract and we
will not be able to check whether the resolution of the problem that
we sketched above does indeed take place. We hope to revisit this in
future work.
\medskip
Before we close the introduction we should mention that the
c.c. problem in the real world is of course much more complicated than
the idealized version that we consider here. Firstly, we only address
one aspect of the c.c. problem which is why the observed value of the
c.c. is not large, that is, not of the order of the cutoff of the
effective field theory and not related questions, such as why the
c.c. has the small positive value that we actually observe, or the
cosmic coincidence problem. Secondly, we have ignored a class of
complications: we study the case of empty, eternal anti-de Sitter
space, while our universe has positive cosmological constant\foot{And
hence the conceptual problems of holography for de Sitter space have
to be addressed.} , has undergone nontrivial cosmological evolution
and is filled with matter. Thirdly, we only consider the fine-tuning
between quantum fluctuations and bare values, while in the real world
the fine-tuning is not only between these two but also additional
contributions to the vacuum energy from the Higgs potential, the QCD
condensate etc. Despite all these differences, we hope that some of
the lessons that we will eventually learn about the c.c. problem in an
idealized anti-de Sitter space may be of relevance to the problem in
the real world.

\newsec{The Main Ideas}

\subsec{\bf The c.c. problem in  the bulk}
\medskip
As we mentioned above, the main goal of this paper is to translate the
cosmological constant problem into a statement in the dual CFT.  A
standard way to express the c.c. problem is to evaluate the 1-loop
vacuum energy of matter fields. In AdS space we have to evaluate the
diagram depicted in figure 1.
\fig{1-loop contribution to the vacuum energy in AdS from a massive fermion.}
{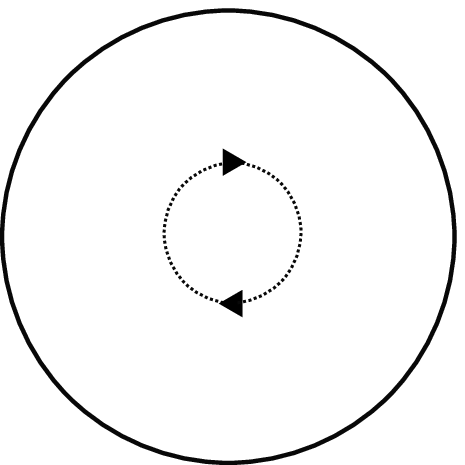}{1.3truein}
\figlabel{\bubblef}
\medskip
We assume that we are in a regime where classical gravity is reliable,
i.e. $R \gg 1/M_P$, where $M_P$ is the Planck mass and $R$ is the AdS
radius. If we call $M_{cut}$ the cutoff of the effective field theory
in the bulk, then this diagram predicts an energy density of the order
$M_{cut}^4$, in the case of four dimensional AdS space. This energy
density enters Einstein's equations multiplied by a factor of $G_N\sim
M_P^{-2}$. Assuming that the cutoff is close to the Planck scale $M_P$
we find that the expected gravitational backreaction of the 1-loop
quantum fluctuations would lead to a modification of the AdS radius to
something of the order $R\sim 1/M_{P}$.
\medskip
If however, and according to our working assumption, what we actually
observe is that $R\gg 1/M_{P}$ it means that somehow the backreaction
of the 1-loop vacuum energy got cancelled. This is usually attributed
to a "counterterm" or equivalently a "bare cosmological constant"
$\Lambda_{bare}$, and the c.c. problem is
that this cancellation between the bare and 1-loop vacuum energies has
to be extremely fine-tuned.
\medskip
What is the interpretation of the diagram in figure \bubblef\ in the
dual CFT? It is not very easy to answer this question, since in
AdS/CFT we are more familiar with calculations of Witten diagrams,
i.e. Feynman diagrams in the bulk where the external points are taken
to infinity in AdS. Luckily, general covariance ensures that the
counterterm responsible for the cancellation of the diagram in
figure \bubblef\ must come in the form

\eqn\maincounter{
\int d^4x \sqrt{-g} \Lambda_{bare}
}
\medskip
\noindent 

Expanding the square root of the metric determinant in \maincounter\
in fluctuations around AdS we find that it generates vertices with
arbitrary number of external gravitons. Indeed, these counterterm
vertices are necessary to (partly) cancel the divergences of the
1-loop diagrams depicted in figure 2.
\fig{1-loop contribution to $n$-point functions of gravitons.}
{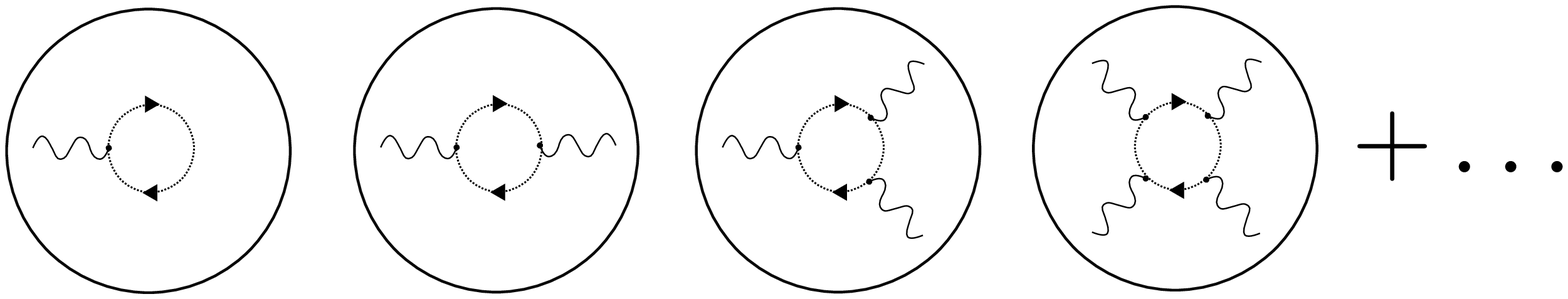}{6.truein}
\figlabel{\bubblecor}
\medskip
Thus the c.c. fine-tuning at the 1-loop level is not only visible in
vacuum-to-vacuum amplitudes but also in $n$-point functions of
gravitons. Let us formulate this more precisely. The smallness of the
bulk cosmological constant is equivalent to the statement that $R\gg
1/M_P$. This in turn means that the connected $n$-point function of
gravitons must be suppressed, relative to the disconnected one, by a
factor\foot{Here we assume that the the energy of the gravitons is
kept fixed in units of the radius of the AdS space, as we take the
$M_P R \rightarrow \infty$ limit. The specific scaling presented here
holds in four bulk dimensions.} of
\eqn\smallcc{
(M_P R)^{2-n}}
in the limit $R\gg 1/M_P$.
\medskip
To be more specific consider, for example, the 4-point function
of gravitons. The first nontrivial contribution to the connected
4-point function comes from the tree-level Born amplitude, which is
indeed of the order $(M_P R)^{-2}$ relative to the disconnected one,
as required by \smallcc.
\fig{1-loop contribution to $4$-point functions of gravitons and c.c.-type counterterm.}
{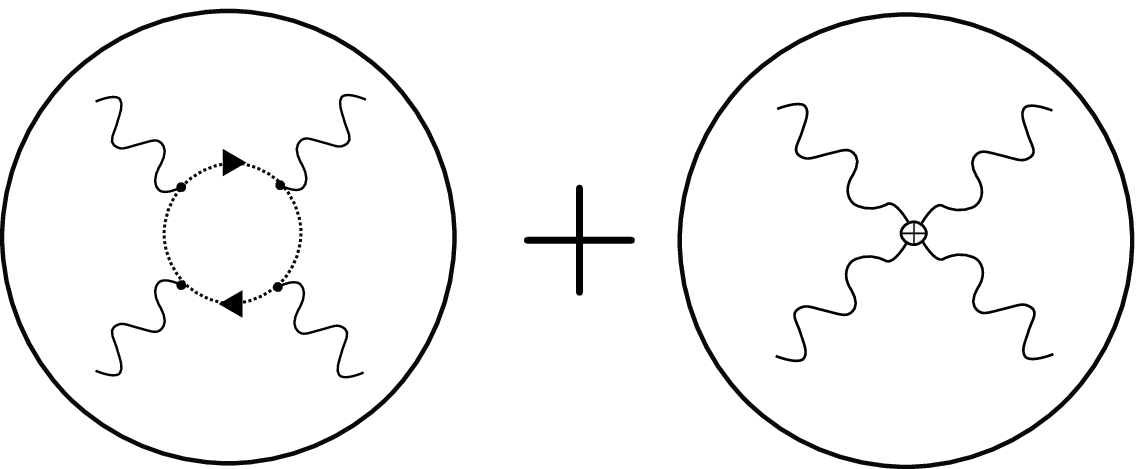}{3.truein}
\figlabel{\fpointcancel}
\medskip
The situation is different however when we look at the 1-loop
correction to the 4-point function due to an internal fermion loop, as
shown in figure 3.  Let us estimate the contribution to this diagram
from the UV regime of the integration over the internal lines. We call
$p$ the momentum running in the loop (for large $p$). Each of the
internal propagators contributes a factor of ${1/p}$ and each vertex a
factor of $p$ from the derivative coupling, which means that the
diagram contains a factor of the form\foot{More details about the UV
divergences of loop diagrams in AdS can be found in later sections.}
$$
\int d^4 p  \sim M_{cut}^4
$$
The diagram is multiplied by a factor of $G_N^2 \sim M_P^{-4}$ from
the vertices. So the contribution of this diagram, relative to the
disconnected one, has an overall factor of
$$
M_P^{-4} M_{cut}^4 
$$
Taking $M_{cut}\sim M_P$ we find that this diagram is ${\cal O}(1)$ in
the $M_P R \gg 1$ limit, and thus violates the requirement \smallcc\
for the scaling of graviton correlators in order for the bulk theory
to have a small c.c. This means that this diagram must be cancelled by
a counterterm to a large extent, as shown in figure \fpointcancel. The
cancellation must be such that the ${\cal O}(1)$ parts completely
cancel giving a sum which is only of the order
$$
(M_P R)^{-2}
$$
To get a feeling of this fine-tuning, if we take the radius of the AdS
space to be of the order of the size of the universe then we have $M_P
R \sim 10^{60}$. Hence the two diagrams depicted in
figure \fpointcancel\ are both of order 1 but have to be cancelled
with each other, leaving behind something of order $10^{-120}$. This
is the analogue of the c.c. fine-tuning expressed in terms of 4-point
functions.
\medskip
Now, 4-point functions of gravitons can be more easily translated into
correlation functions in the CFT, simply by taking the external points
to the conformal boundary of AdS. Then they become Witten diagrams and
via the AdS/CFT prescription can be related to boundary correlators of
the stress energy tensor. Following this strategy, the bulk
fine-tuning can be translated into a statement of fine-tuning purely
within the dual conformal field theory\foot{The reason that we did not
consider the similar story for 0-, 1-, 2- or 3-point functions of
gravitons has to do with the fact that it is more difficult to give
the CFT interpretation of such Witten diagrams, as discussed later.}.

\medskip
In the rest of this paper we will try to develop this point of view.

\subsec{\bf Fine-tuning in the CFT}

Let us now give a short summary of how we think the bulk fine-tuning
manifests itself in the dual CFT. In the previous section we argued
that the fine-tuning can be expressed in terms of graviton correlation
functions. In AdS/CFT correlation functions of gravitons are dual to
correlation functions of the stress energy tensor $T$.  The role of
the parameter $M_P R$ is played by the central charge $c$ \foot{Defined by the 2-point function
of the stress energy tensor.}. If we
redefine the stress tensor $\widetilde{T}=T/\sqrt{c}$ in such a way
that its disconnected correlators are order 1, then the statement that
the observed value of the bulk cosmological constant is small (as
in \smallcc) is dual to the statement that the connected correlators
scale like
\eqn\tscale{
\langle \widetilde{T}(x_1)...\widetilde{T}(x_n) \rangle_{con} \sim c^{(2-n)/2}
} in the $c\gg 1$ limit. In particular we expect that the connected
4-point function will scale like $1/c$. Notice that for large $N$ gauge
theories in the 't Hooft limit we have $c\sim N^2$ and the scaling
mentioned above becomes the usual $1/N$ suppression of connected
correlators.
\medskip
Here is where we start to see the fine-tuning: the bulk discussion of
the previous section suggests that while the 4-point function in the
CFT is of order $1/c$, individual terms which contribute to it may be
parametrically larger and it is only after summing over all of these
contributions that we get the small number of order $1/c$. These
contributions are the CFT dual of the 1-loop and ``counterterm''
Witten diagram depicted in figure \fpointcancel. The question then is, in what
sense can a general CFT correlator be thought of as being the sum of
various terms. Of course we can always artificially split up a correlator into a sum
of terms to make it look fine-tuned. The nontrivial question is whether there is a
canonical way to do so, where these terms have some physical meaning.
\medskip
Indeed, there is a canonical procedure in the CFT which corresponds to
"splitting up" the correlator into a sum of factors. This is the
expansion in conformal blocks
\smallskip
\eqn\cpwexpand{
\langle \widetilde{T}(x_1) \widetilde{T}(x_2) \widetilde{T}(x_3) \widetilde{T}(x_4) \rangle_{con} = \sum_{\cal A} |C_{TT}^{\cal A}|^2 {\bf G}_{\cal A}(x_1,x_2,x_3,x_4)
} From a CFT point of view this is a well-motivated expansion to
consider and moreover is a useful intermediate step before
representing the underlying quantum field theory in a holographic way:
the Witten diagrams in the hologram can be related to certain
(combinations of) conformal blocks with nice properties under crossing
symmetry\foot{See \ElShowkAG\ for a review.}.
\medskip
All of this suggests that the bulk c.c. fine-tuning can be translated
into the statement that while the LHS of equation \cpwexpand\ is of
order $1/c$, there are terms on the RHS of the same equation which are
parametrically larger and which, in the end, cancel among
themselves. If we considered a specific AdS/CFT duality where the bulk
displayed a c.c. fine-tuning problem, and if we were CFT observers, we
would measure a small 4-point function, of order $1/c$, but if we
tried to decompose it into conformal blocks we would be surprised to
find that the smallness $1/c$ is achieved via enormous cancellations
of large terms from individual conformal blocks\foot{The reader might
worry that there may exist some fundamental CFT reason that such a
fine-tuning would not be possible, given bounds from crossing symmetry
and unitarity, for example derived by methods such as those
in \RattazziPE.  However, as we mention later, we do know
(supersymmetric) examples where such fine cancellations between
conformal blocks take place. We thank S. Rychkov for discussions on
these issues.}.

\subsec{\bf A possible resolution of the fine-tuning}

Before presenting more details, let us discuss a speculative
possibility of how holography might provide us with an explanation of
the c.c. fine-tuning, at least in the simplified AdS context that we consider.
\medskip
In AdS/CFT, and presumably more generally in holographic theories, the
degrees of freedom (fields and particles) visible to a low-energy
bulk observer are not related in a simple way to the fundamental
degrees of freedom. In the case of large $N$ gauge theories the
underlying degrees of freedom are quarks and gluons, while the bulk
spacetime describes the dynamics of gauge singlets i.e. glueballs,
mesons, hadrons etc. At the level of effective field theory in the
bulk, an observer only sees the (light) gauge singlets and not the constituent
gluons.
\medskip
An interesting possibility is the following: the mechanism which
guarantees the smallness of the c.c.  (expressed by \tscale) may be
manifest when formulated in terms of the fundamental variables of the
quantum field theory, for example the quarks and gluons, but invisible
in terms of the low-energy fields i.e. the gauge singlets. In such a
case, if we were bulk observers trying to explain the fine-tuning, we
would never be able to understand its origin in terms of the bulk
fields. Only if we gained access to the underlying fundamental theory
(that is, the boundary QFT) would we be able to resolve it.
\medskip
To rephrase, let us imagine that we start with a quantum field theory
(QFT) described by a set of fundamental fields and a path integral
over them. Let us moreover assume that the theory is, or flows to, a
conformal field theory. Using the QFT we calculate the spectrum of
conformal primaries $\Delta_i$ and their 3-point functions
$C_{ij}^k$. We can then describe the QFT in a more abstract way,
simply as a catalogue of the CFT data
\eqn\cftdata{\{\Delta_i, C_{ij}^k\}} Let us call this the CFT language (CFT). If the
CFT has the right properties, i.e. large central charge and a sector
of few low-lying operators whose correlators (approximately)
factorize, then the low-lying sector of the CFT has a natural
representation as a weakly coupled theory in anti-de Sitter space. In
order to describe the bulk theory we only need the CFT data of the
low-lying sector expressed in the form \cftdata. These data translate
into the masses and couplings of what will be the light fields in AdS.
\medskip
\fig{In AdS/CFT we start with a quantum field theory with many fundamental fields. 
Typically there are few operators of low conformal dimensions which
can be described effectively as a perturbation around a
generalized-free-CFT. This effective CFT is naturally represented as a
higher dimensional gravitational theory. The cosmological constant
fine-tuning is visible in the two lower boxes, i.e. the effective CFT
and the gravitational description. It may have a natural explanation
on the deeper microscopic level of the underlying QFT.  }
{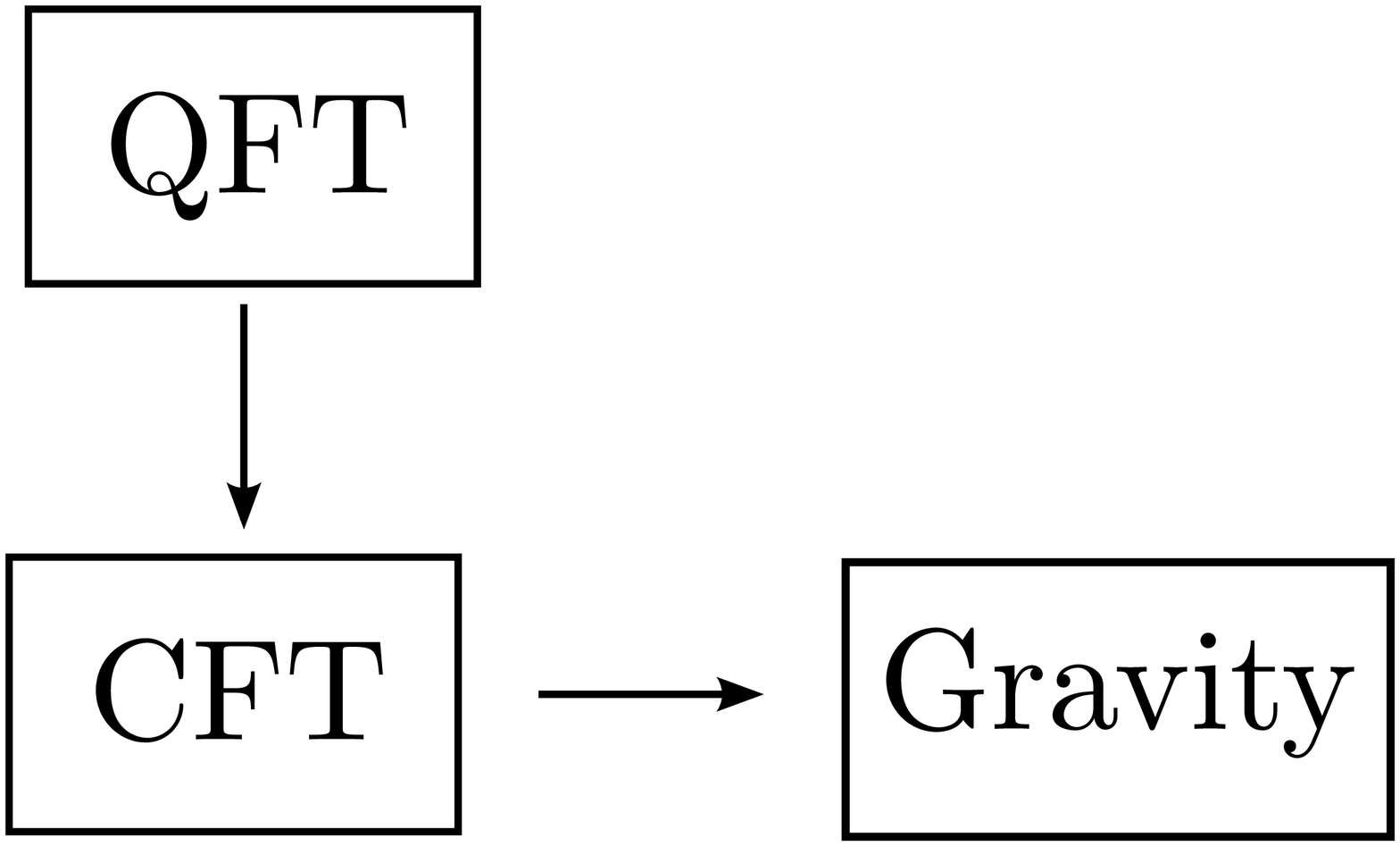}{3.truein}
\figlabel{\trans}

As we will try to explain in this paper, in a theory where the bulk
suffers from fine-tuning, it is precisely the CFT data \cftdata\ that
seem to be ``un-natural''. If we were just given the CFT data, without
the underlying QFT, we would observe large cancellations between
seemingly unrelated terms , for example in double OPE expansions of
the form \cpwexpand, and we would have a hard time explaining
them. These cancellations may have a natural origin in the underlying
QFT due to some dynamical mechanism or symmetry. This mechanism may
become invisible once we pass to the level of the abstract CFT
data \cftdata.
\medskip
This possibility, if true, would suggest that the c.c. fine-tuning between quantum
fluctuations and bare values can be
translated (and resolved) purely within QFT. It can be
translated into a question about the properties of the transformation
that takes us from the upper box of figure \trans\ to the lower left
box. The nontrivial property of this transformation, assuming that our
speculative assumption is true, is that the $1/c$ expansion may look
natural in the first description but fine-tuned in the second\foot{The
two formulations are of course mathematically equivalent. Still, it is
a logical possibility that certain properties i.e. the smallness of
connected correlators may be manifest in one formulation but may look
surprising in the second.}.

\subsec{\bf Comments}

Let us make some clarifying comments:

\bigskip
{\it 1. Is there really a cosmological constant problem in AdS/CFT?}
\medskip
One might think that the existence of a dual CFT somehow removes the
problem altogether, if the CFT is ``stable'', for example if it has
no relevant operators. But this is not the end of the story: while
the existence of a dual CFT may perhaps make the problem look less
dangerous, we would still like to know how the apparent bulk
fine-tuning is resolved. This situation is analogous to the black
hole information problem: while we do believe that the boundary CFT
gives a fully unitary evolution of the system, we still try to
understand how this is achieved in the language of the bulk and what
is wrong with Hawking's argument for information loss\foot{We would
like to thank R. Emparan for pointing out this analogy.}.

\bigskip
\fig{(Left): the hierarchy between the Planck scale and the size of the AdS space $M_P R \gg 1$. (Right): perturbation theory seems fine-tuned if $\xi\equiv M_{cut}^4 M_P^{-2} R^2 \gg 1$.}
{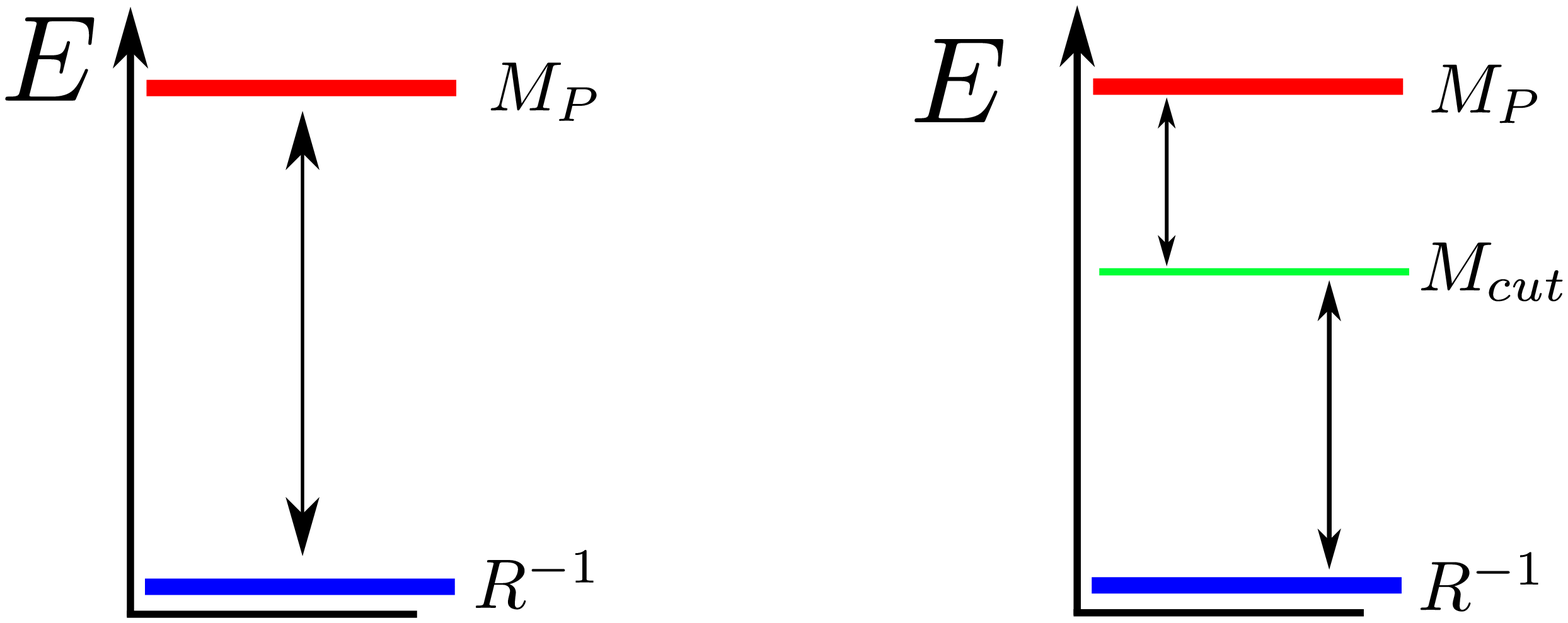}{5.truein}
\figlabel{\disconnected}
\medskip
{\it 2. Hierarchy vs fine-tuning} 
\medskip
Relatedly, one might think that the CFT interpretation of the bulk
cosmological constant problem is simply that the CFT has large central
charge and that this is what creates the large hierarchy. The central
charge controls the parameter $M_P R$, so indeed the large hierarchy
between the Planck and the Hubble scale is mapped to a large value of
the central charge. 
\medskip
The point is that there is an additional puzzle, which is the main
focus of this work, and it has to do with the ``radiative stability''
of this hierarchy from the point of view of perturbation
theory. Besides the scales $M_P$ and $R$ another relevant scale is the
cutoff scale $M_{cut}$ up to which effective field theory in the bulk
is reliable. Unless there is some mechanism (like supersymmetry)
responsible for cancellations, the 1-loop diagrams up to the cutoff
scale contribute a factor of $M_{cut}^D$ to the energy density of the
vacuum\foot{Here by $D$ we refer to the number of macroscopically
large dimensions of the bulk theory. If the CFT is $d$-dimensional it
may be that $D=d+1$, or $D$ could be larger if the duality has a large
internal space, as in the case of IIB on AdS$_5\times$S$^5$. Also, by
$M_P$ we refer to the $D$-dimensional Planck mass.}. Unless there are
nontrivial cancellations with the bare cosmological constant, Einstein
equations predict that the size $R$ of this space should satisfy $
M_{cut}^D M_P^{2-D} R^2\sim 1$.  If, on the other hand we observe that
the combination
$$\xi\equiv M_{cut}^D M_P^{2-D} R^2 \gg 1$$ then it means that some
unexpected cancellations between loop diagrams and counterterms must
be taking place. This fine-tuning is precisely the one we are
interested in.
\medskip
To give an example which shows that there are indeed two separate
aspects of the cosmological constant problem let us compare the
following hypothetical setups: imagine that we have two examples of
AdS/CFT duality, both with very large central charge, both with a
semi-classical gravity dual, up to a very high cutoff scale
$M_{cut}$. Let us moreover assume that example A is
supersymmetric\foot{For example it could be the AdS/CFT duality
between the ABJM theory at large $N$ and M theory on
AdS$_4\times$S$^7$.}, while example B is non-supersymmetric. In both
cases the combination $M_P R$ is very large (since the central charge
is large) and bulk observers could legitimately ask why they happen to
live in a world with such large central charge.
\medskip
The difference is that in case A the second aspect of the puzzle,
which has to do with radiative corrections, is absent. It is clear how
1-loop diagrams are cancelled among themselves due to
supersymmetry. On the other hand in case B, the bulk observer would
wonder, not only why the central charge is large, but also how the
fine cancellations between loop-diagrams and bare values (or
counterterms) are achieved.
\medskip
In other words: if, hypothetically, in our world we found that
supersymmetry was restored at a very low scale, say of the order of
$\sim 10^{-3} eV$, then the cosmological constant problem would become
less puzzling. We would perhaps still try to explain the origin of the
large hierarchy between the Hubble and the Planck scale, but there
would be no puzzle about how the loop diagrams cancel against
counterterms: a mechanism, cancellation between bosons and fermions,
would guarantee the radiative stability.
\bigskip
{\it 3. The Role of Supersymmetry} 
\medskip
From the above it should be clear that we are only interested in
non-supersymmetric examples of the AdS/CFT correspondence. In
supersymmetric examples we know that bosonic and fermionic loop
diagrams cancel each other. In AdS space there may be a small mismatch
between the radiative corrections of bosons and fermions\foot{This can
also be understood from the CFT point of view: in representations of
the superconformal algebra bosonic and fermionic states do not have
the same energy, as the supersymmetry operators do not commute with the
Hamiltonian on the sphere (i.e. the dilatation operator).}, but this
at most of the order of the AdS scale $1/R$. At higher energy scales,
where we can ignore the curvature of AdS space, supersymmetry does
indeed lead to cancellations between loop diagrams.  Hence loop
diagrams may give non-trivial contributions only up to the scale
$M_{SUSY}\sim 1/ R$ and we have, in the language of the previous
section, that the relevant ratio $\xi$ is
$$
\xi \sim M_{SUSY}^D M_P^{2-D} R^2 \sim (M_P R)^{2-D} \ll 1
$$
So there is no need for any other cancellations in perturbation
theory, besides those between bosonic and fermionic loops.
\smallskip
Notice however that the cancellation between bosonic and fermionic
loop diagrams means that if we consider the double OPE
expansion \cpwexpand\ for a supersymmetric CFT with a bulk cutoff near
the Planck scale we would observe the following fine-tuning: the LHS
would be of order $1/c$, but on the RHS we would have (besides other
terms) partial sums over double trace operators of bosons and
fermions, where each of these sums is parametrically larger than $1/c$
but with opposite signs. So even in the supersymmetric case, there are
large cancellations in the conformal block expansion but in this case
the mechanism responsible for these cancellations is a well understood
one, namely supersymmetry.
\medskip
We do not want to discuss the supersymmetric examples any further. It
is well known that supersymmetry might offer an explanation of the
c.c. fine-tuning, but this would require a very low scale of
supersymmetry breaking, which is experimentally ruled out.

\bigskip
{\it 4. The string scale and other cutoffs} 
\medskip
In many examples of AdS/CFT there is a ``string scale'' $M_s$, besides
the Planck scale $M_P$. Typically in the large $N$ limit the string
scale is of order $f/R$, where $f$ is a factor which depends on other
parameters, such as the 't Hooft coupling. No matter how large $f$ is,
in the conventional large $N$ limit it does not scale with $N$. So in
the large $N$ limit we have $M_P/M_s \rightarrow \infty$. Independent
of the existence of supersymmetry, the presence of such a string
scale, parametrically smaller than the Planck scale, removes the sharp
fine-tuning problem. The loop diagrams can be trusted only up to the
string scale hence $M_{cut}\sim {f \over R}$ and we find
$$
\xi \sim {f^D \over R^D} M_P^{2-D} R^2\ll 1
$$
which means that while the energy density due to loop diagrams is of
order $M_S^D$, it is not heavy enough to backreact in the limit $M_P
R\rightarrow \infty$.
\medskip
Notice that even in bosonic string theory in flat space, ignoring the
tachyon instabilities for a moment, the 1-loop vacuum diagram gives a
contribution to the c.c. of order $M_s^D$ which is parametrically
smaller than $M_P^D$ if $g_s\ll 1$.
\medskip
So the (well-known) conclusion is that a low string scale would
resolve\foot{Or more accurately it prevents us from posing a
sharp paradox.}  the c.c. fine-tuning problem by effectively lowering the
cutoff of the effective field theory. But this resolution is not
suitable for us, since it would require, in the real world, an
unrealistically low string scale (i.e. of the order of $M_{cut}\sim 10^{-30} M_P\sim10^{-3}eV$).
\medskip
From the AdS/CFT point of view these observations show that  large $N$ gauge theories in the 't Hooft limit may be
less suitable examples/toy models to study the problem. These theories
have duals with a string scale parametrically lower than the Planck
scale, as can be seen from the Hagedorn growth in the spectrum of
single trace operators \PolyakovAF, \SundborgUE, \AharonySX. Hence the
1-loop vacuum energy up to the cutoff of effective field theory is not
expected to backreact significantly and we cannot pose a sharp
c.c. problem in the bulk.
\smallskip 
The same problem exists for bosonic examples of AdS/CFT where the bulk
is described by some higher spin theories in AdS, such
as \KlebanovJA, \GaberdielPZ, \KiritsisXC. As for the stringy
examples, in these models the effective field theory in the bulk has a
very low cutoff, of the order of the AdS scale, so we have that the
ratio $\xi\ll 1$ and we cannot pose a sharp c.c. problem.

\bigskip

\newsec{CFTs with holographic duals}

In order to translate the bulk fine-tuning problem into a statement in
the dual CFT we first have to review the basic properties of CFTs with
a holographic description.  We try to be brief and refer the
interested reader to \ElShowkAG\ for more details.
\medskip
The most characteristic properties of such CFTs is that they have
large central charge $c$ and that their spectrum at low conformal
dimensions contains a sector with a small number of ``generalized free
fields'', i.e. operators whose correlators factorize \ElShowkAG. The
factorization has to be understood as a statement which holds in the
large $c$ limit, i.e. up to $1/c$ corrections. This sector of
operators can be naturally represented in terms of a higher
dimensional gravitational theory in AdS space. In general this
gravitational theory will be very complicated and not well described
by Einstein gravity.  For example, it may be a highly curved (though
weakly coupled) string theory, or higher spin gravity.
\medskip
One final condition, on top of the ones described in the previous
paragraph, guarantees that the bulk theory simplifies significantly
and can be described by semi-classical two-derivative gravity: the
condition that single-particle operators with spin higher than 2 have
conformal dimension parametrically larger than that of the stress
energy tensor, as emphasized in
\HeemskerkPN. In general let us call $\Delta_{cut}$ the conformal dimension where the higher spin single-particle operators appear. Semi-classical gravity becomes a good approximation if $\Delta_{cut}\gg 1$.

\subsec{\bf Candidate CFTs with c.c. problem}

According to the previous discussions a CFT whose bulk dual displays
the c.c. problem must have the following properties:
\bigskip
i) In the first place it must satisfy all the conditions of a
holographic CFT i.e. large central charge, few light operators,
factorization etc.
\medskip
ii) The CFT should not be supersymmetric\foot{Otherwise the
cancellation of the zero-point energy is well understood and there is
no puzzle to be explained.}.
\medskip
iii) The cutoff of the effective theory in the bulk should be high enough so
that the 1-loop vacuum energy can backreact significantly
i.e. $M_{cut}^{D} M_P^{2-D}R^2 \gg 1$. Translating this into a
statement about the conformal dimension $\Delta_{cut}$ where higher
spin/stringy states start to appear we find the condition
\eqn\condfine{
\Delta_{cut}\gg c^{1/D}
}
\smallskip
At the moment we do not know any CFT satisfying all these
conditions. Examples of CFTs satisfying conditions i) and ii), but not
iii), are non-supersymmetric orbifolds of supersymmetric string backgrounds 
such as \KachruYS, the large $N$ $O(N)$ models in 3 dimensions or the models
discussed in \GaberdielPZ, \KiritsisXC. These theories do not satisfy
iii) because the cutoff of effective field theory in the bulk is of
the order of the AdS scale, or $\Delta_{cut} \sim {\cal O}(1)$ and
not \condfine\foot{Because the higher spin fields
appear at conformal dimension of order 1.}.
\medskip
On the other hand an example of a CFT satisfying i)
and iii), but not ii), is the ABJM theory at large $N$ (and
$k=1$). While the spectrum of operators in this CFT cannot be studied
directly from the gauge theory side, we expect from the bulk that the
only UV scale is the $D=11$ dimensional Planck scale, hence we expect
$\Delta_{cut}$ to satisfy \condfine\foot{Since M-theory is only
characterized by the $D=11$-dimensional $M_P$ (there is no "string
scale") we expect that at large $N$, the new operators going beyond
11d supergravity will appear at energies of order $M_P$. Translating
into conformal dimensions it means $\Delta_{cut} \sim c^{1/(D-2)}$
which is sufficient for \condfine.}.
\medskip
It would clearly be fascinating to find a CFT where all of these
properties i), ii), iii) are simultaneously satisfied. Such a CFT would describe
holographically AdS gravity without supersymmetry. Even though such a
CFT is not yet known, we will assume that it exists and, guided by the
bulk description, try to understand how the c.c. fine-tuning would
manifest itself in abstract CFT language.

\subsec{\bf Witten diagrams and large $c$ expansion} 

As we explained in the introduction the bulk c.c. problem can also be
phrased in terms of Witten diagrams in the bulk. Hence in order to
translate the problem in CFT terms it is crucial to understand the
meaning of the Witten diagram expansion from the CFT point of view.
\medskip
This is a general question, with other possible applications, but
which has never been fully clarified, especially to higher orders in
$1/c$, when loop Witten diagrams become important. In any case, we
review what is known so far and we hope the CFT interpretation of the
Witten diagram expansion will be understood better in the future.
\medskip
As argued in \ElShowkAG\ and also in \HeemskerkPN, \HeemskerkTY, \FitzpatrickZM, the
AdS space can be thought of as a convenient way to represent the $1/c$
expansion around a generalized free CFT. The Witten diagram expansion
around a generalized free CFT play a similar role as the Feynman
diagram expansion around a standard free field theory.
\medskip
\fig{Witten diagrams and conformal blocks}
{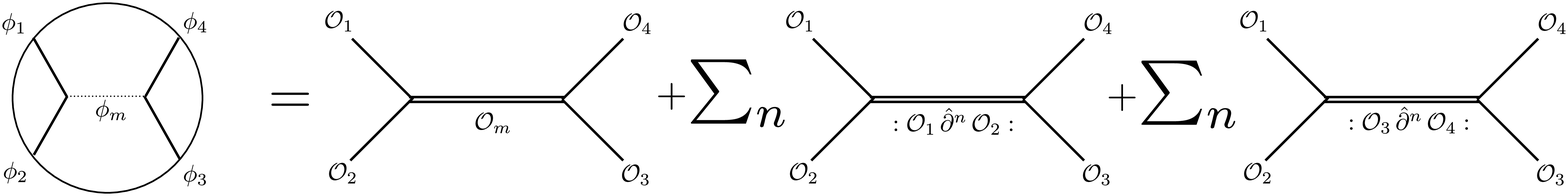}{6.truein}
\figlabel{\wittencpw}
\medskip
In figure \wittencpw\ we see a basic scalar exchange Witten
diagram. At first one might expect that this Witten diagram
corresponds in the CFT to the conformal block of the operator ${\cal
O}_m$ dual to the exchanged field $\phi_m$ in the bulk. This is almost, but not
quite correct. Besides the conformal block of ${\cal O}_m$ the
exchange Witten diagram contains an infinite sum over
two-particle\foot{Or ``double trace'' in the language of large $N$
gauge theories.} operators \LiuTH, as shown in the figure. Why are these
operators there and what fixes their relative coefficients? The answer
is that in theories with a large $c$ expansion there is a unique way
that the basic conformal block ${\cal O}_m$ can be ``dressed up'' by
two-particle conformal blocks into a combination which has a good OPE
expansion in the crossed channel
\ElShowkAG. This combination is precisely the exhange Witten diagram depicted above. 
\medskip
This correspondence can be generalized to other kinds of Witten
diagrams. More specifically, in order to understand how to deal with
loop diagrams we have to address the issue of choice of basis of
operators. When considering the large $c$ expansion of CFTs we are not
looking at one and fixed CFT but rather at a family of CFTs
parametrized by $c$. Hence we have a double expansion: we have an expansion 
in terms of powers of $1/c$ and an expansion in terms of conformal blocks. In this
situation we have to understand how we choose the basis of operators
as a function of $c$. A natural choice is to work in terms of the
``instantaneous'' conformal primaries ${\cal O}_i(c)$. In this basis
the 2- and 3-point functions are fixed by conformal invariance and
some of the Witten diagrams, such as the self-energy diagrams, are
more difficult to interpret.
\medskip
Another choice of basis, which seems to be closer to what the Witten
diagram expansion represents and which was nicely discussed
in \FitzpatrickZM\ and also \FitzpatrickHH, \FanWM, is to try to build
the eigenstates (conformal primaries) of the interacting theory by
using the Hilbert space of the free one. This is the approach that we
usually follow in perturbation theory in quantum mechanics where the
interacting eigenstates are written as superpositions of the free
ones. From this point of view we start with a free Hamiltonian $H_0$
describing the propagation of free fields in AdS. This is dual to the
dilatation operator in the $c=\infty$ conformal field theory. We then
perform perturbation theory in the small parameter $g\sim
1/\sqrt{c}$. The interacting Hamiltonian/dilatation operator has the
form
\medskip
\eqn\pertham{
H = H_0 + g V_1 + g^2 V_2 +\dots
}\medskip
\noindent These operators are constructed to act on the Hilbert space of the generalized free fields 
i.e. the (low-lying) gauge singlets.
\medskip
The eigenstates (conformal primaries) of the interacting 
Hamiltonian $H$ can be written as superpositions of the eigenstates of
the free theory $H_0$ by doing standard quantum mechanics perturbation
theory. This leads to mixing between single- and multi-particle
operators. This perturbation theory is related to the Witten diagram
expansion, though clearly the details have to be worked out more
carefully.

\subsec{\bf Radiative corrections in AdS/CFT}

In most examples of AdS/CFT that we know, the parameter which controls
the strength of the bulk interactions is $1/N$ (or more generally the
inverse central charge $1/c$). This parameter is related to the combination
$M_P R$. This means that once we start considering loop diagrams, we
inevitably also have to consider quantum gravity effects. A related
issue is that in many examples the cutoff of the effective
(super)-gravity theory is of the order $M_P$ i.e.  it is correlated to
the loop-expansion parameter.
\medskip
Both of these issues can make things a little more complicated. So we
will start by first taking the central charge (i.e. the combination
$M_P R$), the loop-counting parameter $g$ and the cutoff $M_{cut}$ to
be (formally)  independent quantities. We will consider loop
diagrams to determine how they depend on
these parameters. In the end, if we want to apply our results to a
particular example of AdS/CFT, we have to go back and see how all
these quantities are correlated.
\medskip
In the rest of this short section we will discuss how we can introduce
a UV cutoff in an AdS covariant way to regulate the loop momentum
integrals and what is its meaning in the dual CFT.
\bigskip
\noindent {\it Cutoff in AdS}
\medskip
Let us imagine that we are observers in AdS and that we try to
describe physics by some sort of effective field theory. For this we
need to introduce a UV cutoff $M_{cut}$ to regulate the divergences of
loop integrals and to parametrize the regime of validity of our
efffective description.
\medskip
When doing quantum field theory in curved space it is important to
impose the cutoff in a covariant way. In flat space we can think of a
cutoff as a maximum energy/momentum allowed to run in the loops. In
AdS space there is a global notion of energy, but this cannot be used
as a cutoff, due to the large redshift factor between the ``center''
and the ``boundary'' of AdS. Instead the cutoff has to be imposed on
the energy of virtual particles as measured in a local inertial frame\foot{We would
like to thank S. Minwalla for discussions on these issues.}.
\medskip
In practice if we use a Pauli-Villars regularization method, it will
automatically introduce a correct covariant cutoff.
\bigskip
\noindent{\it Cutoff in the CFT}
\medskip
At first one might think that the bulk cutoff translates in the CFT
into a cutoff in the conformal dimensions of the form
$\Delta_{cut} \sim M_{cut} R$. However, as we explained above, this is
not a covariant cutoff and is in fact inconsistent with conformal
invariance. By acting with the ``raising operators'' $P_\mu$ we can
indefinitely increase the conformal dimension of an operator. So it is
not consistent to impose a sharp conformal dimension cutoff.
\medskip
Instead what we have to do is to impose a cutoff on the ``relative
momentum'' of many-particle states \FitzpatrickZM. This can be done as
follows. States with single particles are related to single trace
operators ${\cal O}$. States with two particles are related to
``double-trace operators''. When such operators are conformal
primaries they have the form
$$
:{\cal O} (\partial^2)^n \partial_{[1}...\partial_{l]} {\cal O}:
$$
At infinite $c$ this 2-particle state has conformal dimension $2\Delta+2n+l$. It
represents two particles, whose ``center of mass'' is at rest in the
center of AdS and the relative momentum is controlled by $n,l$. By
considering the descendants of this bound state, we can set the center
of mass in motion in AdS.
\medskip
It is then clear that the cutoff has to be imposed in the CFT by
bounding the parameters $n,l$ in the conformal primaries of
two-particle operators (and similarly for three- and higher- particle
operators). This type of cutoff is consistent with conformal
invariance, or equivalently, covariant with the curved metric of AdS.

\newsec{A simple example of a 1-loop diagram}

In this section we consider a simple 1-loop diagram in AdS/CFT and try
to find its interpretation in the boundary CFT. We are especially
interested in the meaning of the UV divergences and of the
counterterms introduced to cancel them.

\subsec{\bf Some background} 

We consider the AdS/CFT correspondence where we have a $d$-dimensional
CFT and a $D=d+1$ dimensional AdS space. In AdS/CFT single-particle
operators ${\cal O}$ of the CFT are associated to fields propagating
in the bulk. For scalar operators we have the relation
\eqn\massdim{
\Delta = d/2 +\sqrt{m^2 + (d/2)^2}
} between the conformal dimension $\Delta$ of ${\cal O}$ and the mass
$m$ of the dual bulk field $\phi$. Correlation functions of ${\cal O}$
can be computed in (bulk) perturbation theory by evaluating Witten
diagrams. The bulk-to-boundary propagator for a scalar field of mass
$m$ has the form
$$
K_m(x,z) ={\Gamma(\Delta)\over \pi^{d/2} \Gamma(\Delta-d/2)}
\left({z_0 \over z_0^2 +(\vec{z}-\vec{x})^2}\right)^\Delta 
$$
where the vector $\vec{x}$ denotes a point on the conformal boundary
of AdS and $z=(z_0,\vec{z})$ a point in the bulk in (Euclidean)
Poincare coordinates where the metric is
$$
ds^2 = {dz_0^2 + d\vec{z}^2 \over z_0^2}
$$
We also need the bulk-to-bulk propagator
$$
G_m(z,w) = {\Gamma(\Delta) \over 2^{\Delta+1}\pi^{d/2} \Gamma\left(\Delta-{d-2\over 2}\right)} \,s^\Delta\,
_2 F_1 \left({\Delta \over 2}, {\Delta+1 \over 2}, \Delta -{d-2\over 2},s^2 \right)
$$
where
$$
s= {2 z_0 w_0 \over z_0^2 + w_0^2 + (\vec{z}-\vec{w})^2}
$$
Internal vertices in the bulk have to be integrated over their
position in AdS with the measure
$$\int {d^D z \over z_0^D }$$

\subsec{\bf Bulk}

Let us see how we can compute a simple 1-loop diagram in AdS. We
 consider the $\phi^4$ theory in the bulk
 $$
{\cal L} = {(\nabla\phi)^2\over 2}  +{m^2 \phi^2 \over 2} +{g\over 4!} \phi^4
$$ 
and for the moment we work in general dimensionality of AdS$_{D}$
where, depending on $D$, the interaction can be relevant ($D<4$),
marginally irrelevant $(D=4)$ or irrelevant $(D>4)$.  The coupling
constant $g$ has dimensions $[{\rm mass}]^{4-D}$.  We will study the
Witten diagrams contributing to the 4-point function of the operator
${\cal O}$ dual to the bulk field $\phi$ in perturbation theory in
$g$. We can write
$$
\langle {\cal O}(x_1)  {\cal O}(x_2)  {\cal O}(x_3)   {\cal O}(x_4) \rangle
= \sum_{n=0}^\infty W^{(n)}(x_1,x_2,x_3,x_4)
$$
where the contribution $W^{(n)}$ is of order $g^n$ and, as usual in
quantum field theory, the equality has to be understood as an
asymptotic expansion.
\fig{Disconnected Witten diagrams.}
{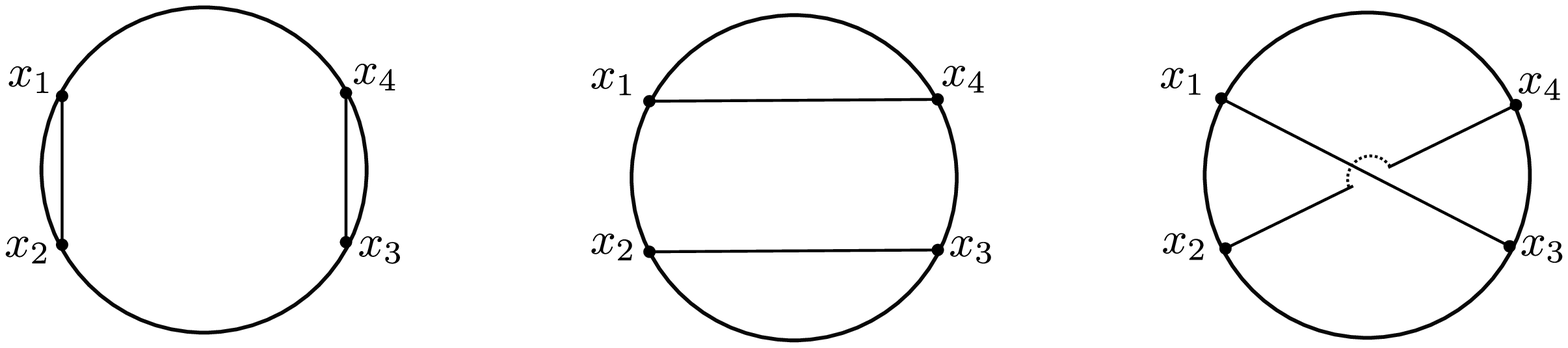}{5.truein}
\figlabel{\disconnected}
At zeroth order in $g$ we only have the disconnected components
depicted in figure \disconnected. They have a simple factorized form
\eqn\discon{
W^{(0)} \sim {1\over |x_1-x_2|^{2\Delta}|x_3-x_4|^{2\Delta}} + {\rm permutations}
}
\fig{Tree level Witten diagram.}
{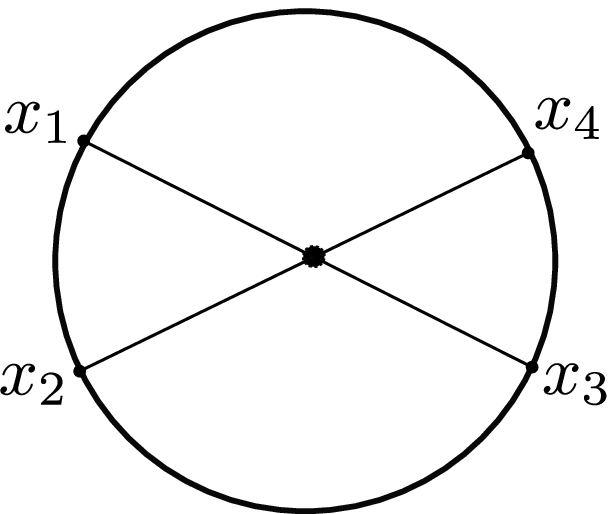}{1.3truein}
\figlabel{\tree}
\medskip
At order $g$ we have to evaluate the tree level diagram of
figure \tree. This diagram is equal to
$$
W^{(1)}   \sim g \,{\bf D}_m(x_1,x_2,x_3,x_4)\equiv g \int {d^D z \over z_0^D} K_m(x_1,z) K_m(x_2,z) K_m(x_3,z) K_m(x_4,z) 
$$
where we have introduced the ``${\bf D}$-function'' for a field of
mass $m$.
\fig{1-loop Witten diagrams.}
{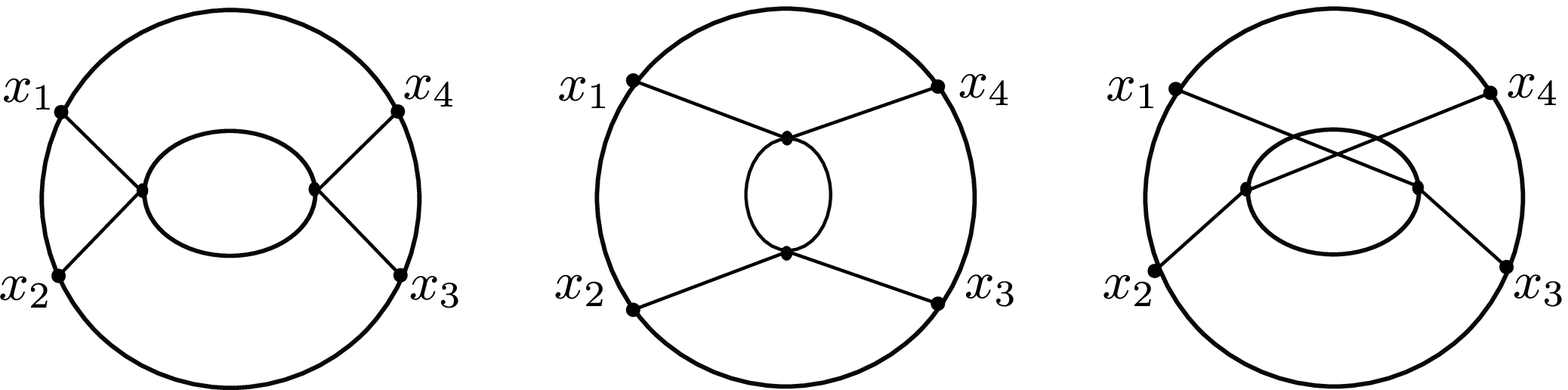}{4.5truein}
\figlabel{\loopa}
\medskip
At order $g^2$ we have the loop diagrams depicted in figure \loopa\
which are equal to\foot{We also have loop diagrams with corrections to
the propagators of the external legs. Such diagrams are related to
corrections of the dimension of the operator ${\cal O}$. For
simplicity we ignore them in this section.}
$$
W^{(2)}(x_1,x_2,x_3,x_4) \sim {g^2\over 2}  \int {d^D z \over z_0^D}
 {d^D w \over w_0^D} K_m(x_1,z) K_m(x_2,z) G_m(z,w) G_m(z,w) K_m(x_3,w) K_m(x_4,w) 
$$
$$
+ {\rm permutations}
$$
The UV behavior of these integrals can be estimated from the
equivalent diagrams in flat space. If we call $p$ the momentum running
in the loop, then the two internal propagators contribute a factor of
${1\over p^4}$, at large $p$, while the integration over momenta a
factor of $d^Dp$. The degree of divergence of the diagram is $D-4$.
Hence we expect this integral to be UV divergent for $D>4$,
logarithmically UV divergent for $D=4$ and finite for $D<4$.  Since
these estimates have to do with a UV divergence we expect them to be
the same in AdS space.
\medskip
We can check this in the following way: the divergence is coming from
the region $z\sim w$. For the moment let us fix $w$ and expand $z$ as
$z=w+w_0 \epsilon$, where $\epsilon=(\epsilon_0,\vec{\epsilon})$ is a
small $D$-dimensional coordinate vector in Poincare coordinates. The
distance $s$ can be expanded as
$$
s = 1 - \epsilon^2 +...
$$
where $\epsilon^2 = \epsilon_0^2 + \vec{\epsilon}\,^2$ and then the
bulk-to-bulk propagator has the following short-distance expansion
$$
G_m(w+\epsilon,w) \sim |\epsilon|^{-(D-2)}+ {\rm subleading}
$$
This is indeed the expected leading short-distance behavior of the propagator of a scalar field in flat space. 
\medskip
Considering the integration over $z$ in a small neighborhood around
$w$ (say down to $\epsilon\sim {1\over M_{cut}}$) we find that the
product of two propagators times the measure from $d^D z$ has a UV
divergence of the form $M_{cut}^{D-4}$, for $D> 4$, and $\log M_{cut}$
for $D=4$. Hence, in the case $D<4$ the integral is finite and can in
principle be computed in straightforward way. In $D\geq 4$ we need to
regularize and renormalize.

\subsec{\bf Renormalization and counterterms in $D=4$.}

We now review how this can be done in the case $D=4$, similar analysis
can be performed in higher dimensions. We find it practical to adopt a
Pauli-Villars regularization method. We introduce a fictitious heavy
field $\Phi$ of mass $M_{PV}$ and wrong-sign kinetic term. Hence we
consider the more general regularized action:
$$
{\cal L}_{reg} = {(\nabla\phi)^2 \over 2} +{m^2 \phi^2 \over 2} 
- {(\nabla\Phi)^2 \over 2} -{M_{PV}^2 \Phi^2 \over 2} + {g\over 4!} (\phi+\Phi)^4
$$
The bulk-to-bulk propagator of the field $\Phi$ is the opposite of
that of a field with mass $M_{PV}$. In practice, given a loop diagram
with external $\phi$ lines, we simply have to replace the internal
bulk-to-bulk propagators by
$$
G_m (z,w) \rightarrow G_m(z,w) - G_{M_{PV}}(z,w)
$$
In this way we can define the regularized 4-point function as
$$
W^{reg}_{1-loop}(x_1,x_2,x_3,x_4;M_{PV})\sim{g^2\over 2} \int {d^4
 z \over z_0^4} {d^4 w \over w_0^4}\times
$$
$$
\times K_m(x_1,z) K_m(x_2,z) \left( G_m(z,w) - G_{M_{PV}}(z,w)\right) \left(G_m(z,w) - G_{M_{PV}}(z,w)\right)K_m(x_3,w) K_m(x_4,w) 
$$
$$
+ {\rm permutations}
$$
For finite $M_{PV}$ this integral is convergent. In the limit
$M_{PV}\rightarrow\infty$ it diverges logarithmically, as we expect
from flat space. The Pauli-Villars mass $M_{PV}$ can be qualitatively
identified with the UV cutoff $M_{cut}$.

In order to define ``renormalized'' correlators we have to introduce a
counterterm in the Lagrangian, chosen so that the sum of the 1-loop
diagram and the counterterm is finite as we send the cutoff to
infinity. For the diagram under consideration this can be achieved by
adding
$$
{\cal L}_{c.t.} ={ \delta_g \over 4!} \phi^4
$$
where $\delta_g$ has to be chosen to cancel the divergent part of the
1-loop integral. This happens if we take
$$
\delta_g= - b {g^2\over 16 \pi} \log \left({M_{PV} \over \mu} \right)
$$
(to fix the overall numerical constant $b$ we have to keep track of
all factors in the intermediate formulas).  The scale $\mu$, which
corresponds to a finite shift of the counterterm, can be fixed by a
renormalization condition.

\fig{1-loop counterterm.}
{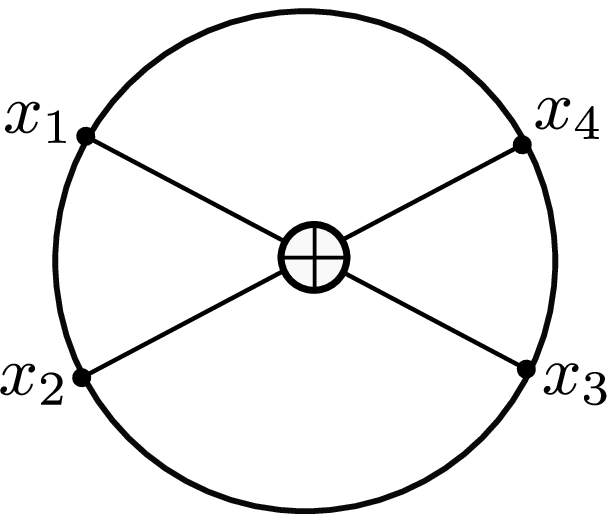}{1.5truein}
\figlabel{\countert}
\medskip
Putting everything together we find that the 1-loop diagram together
with the counterterm give a contribution
$$
W^{reg}(x_1,x_2,x_3,x_4;M_{PV}) \sim W^{reg}_{1-loop}(x_1,x_2,x_3,x_4;M_{PV}) - b {g^2\over
16\pi} \log\left({M_{PV}\over \mu}\right) {\bf D}_m(x_1,x_2,x_3,x_4)
$$
The ``renormalized'' 4-point function at order $g^2$ is computed by
sending the regulator cutoff to infinity
$$
W^{(2)}(x_1,x_2,x_3,x_4)=\lim_{M_{PV}\rightarrow \infty}W^{reg}(x_1,x_2,x_3,x_4;M_{PV}) 
$$
The result is finite and unambiguous up to the choice of the finite
term $\mu$, which can be fixed by choosing an appropriate
renormalization condition/i.e. redefinition of the coupling constant
$g$.

\subsec{\bf Boundary interpretation and fine-tuning}

Let us now understand the meaning of these diagrams in the dual
CFT. As we mentioned, the interpretation of Witten diagrams in the CFT
has never been fully clarified, but it is known that a CFT notion that
comes close to a Witten diagram expansion is the expansion of
correlators in conformal blocks, see \ElShowkAG\ for more details. In
a CFT it is natural to analyze a 4-point function by expanding it in a
double OPE, say in the $(12)\rightarrow (34)$ channel, as
\eqn\doubleope{
\langle {\cal O}(x_1) {\cal O}(x_2) {\cal O}(x_3) {\cal O}(x_4)\rangle
=\sum_{\cal A} |C_{{\cal O}{\cal O}}^{\cal{A}}|^2 {\bf G}_{\cal
A}^{12,34} (x_1,x_2,x_3,x_4) } where the sum runs over all conformal
primary operators ${\cal A}$. The coefficients $C_{{\cal O}{\cal
O}}^{\cal A}$ depend on the dynamics of the CFT, while ${\bf G}_{\cal
A}^{12,34} (x_1,x_2,x_3,x_4)$ are the conformal blocks i.e. special
functions whose form is fixed by kinematics of the conformal
group \DolanUT.  In theories with weakly coupled holographic duals the
Witten diagrams encode certain combinations of conformal blocks with
simple behavior under crossing symmetry.
\fig{Expansion of Witten diagrams in conformal blocks.}
{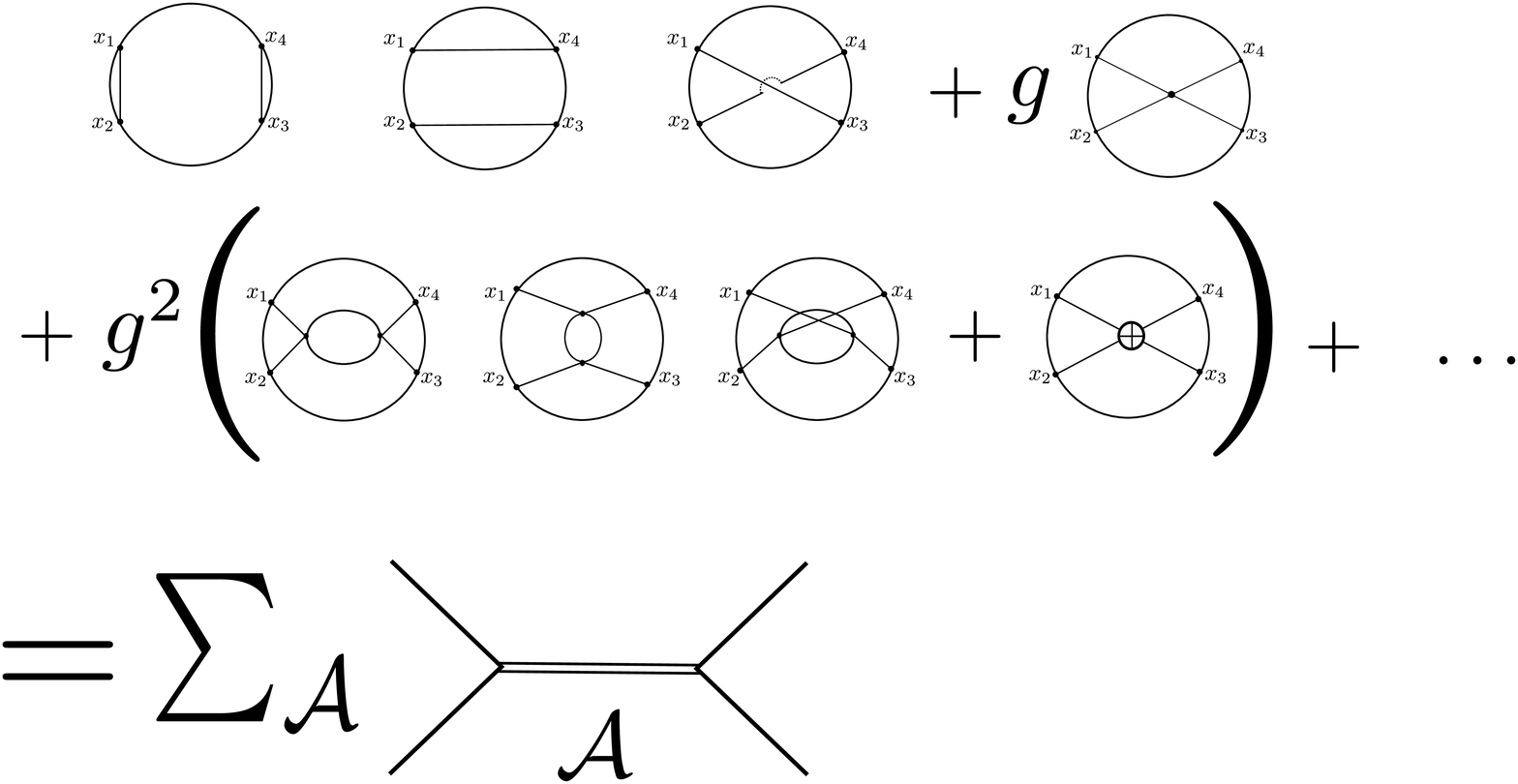}{4.5truein}
\figlabel{\wittencpw}
\medskip
Without going into details, let us illustrate this by first explaining
the meaning of the disconnected and tree-level interactions.
\medskip
When $g=0$, the correlators factorize to products of 2-point
functions. If we take the factorized correlator
$W^{(0)}(x_1,x_2,x_3,x_4)$ given in \discon\ and perform an expansion
of the form \doubleope\ we find that what runs in-between is the
identity operator $I$, as well as two-particle conformal primary
operators of the form $:{\cal O}
(\partial^2)^n \partial_1...\partial_l {\cal O}:$.  All these
operators come with coefficients of order one, which can be easily
computed \DolanUT.
\bigskip
The tree level diagram $W^{(1)}$ can also be expanded in conformal
blocks and we find that it corresponds to the exchange of two-particle
operators of the form $:{\cal O}(\partial^2)^n{\cal O}:$ with
coefficients of order $g$. As shown in \HeemskerkPN\ this is the only
combination of such (scalar) operators which is consistent with
crossing symmetry.
\bigskip
Let us now consider the 1-loop diagram $W^{(2)}$, first in the case
$D<4$ where the diagram is convergent by itself, without any
counterterms. In this case it was nicely shown in \PenedonesUE, using
the Mellin representation of CFT correlators, that the loop diagram
corresponds to the exchange of two-particle operators of the form\foot{Notice that
in this case the exchanged double trace operators are the same for all of these
diagrams. If we had a field ${\cal O}'$ running in the intermediate loop then the
exchanged double trace operators form the loop diagram would be of the form $:{\cal O}'
(\partial^2)^n \partial_1...\partial_l {\cal O}'$.}
\eqn\doubleloop{  
:{\cal O} (\partial^2)^n \partial_1...\partial_l {\cal O}: } These
operators come with coefficients of order $g^2$. This is intuitively
reasonable, cutting the diagram along the loop we see a two-particle
state propagating in the middle. The number of derivatives in the
conformal primary \doubleloop\ encodes the momentum running in the
loop.
\medskip
The integral over the momentum in the bulk translates into an infinite
sum over two-particle conformal primaries of this form, with increasing
number of derivatives. As we explained, in the case $D<4$ this sum is
convergent.
\medskip
In the more interesting case of $D=4$ we expect a (logarithmic) UV
divergence\foot{Of course this sum will be divergent for any $D>4$, we consider $D=4$ for 
simplicity}. We first interpret the diagram with a cutoff
$M_{cut}\approx M_{PV}$. Expanding it in conformal blocks we find that
(up to the cutoff) it corresponds to the exchange of double
trace-operators of the same form, with coefficients of order
$g^2$. However now we notice something very interesting: while each of
the contributions of a conformal block is of order $g^2$ (i.e. small),
the sum over conformal blocks of two-particle operators gives a
contribution which diverges like $g^2 \log M_{cut}$.
\medskip
In the bulk we corrected this divergence by introducing a
``counterterm'', i.e. a diagram with the same structure as the tree
level contact interaction, but with a coefficient which is large, of
order $g^2 \log M_{cut}$ (in order to cancel the divergent 1-loop
diagram). In the double OPE expansion this means that some of the
double trace operators will have to come with very large coefficients
in order to cancel the large contribution encountered above from the
(partial) sum over conformal partial waves from the loop diagram.
\bigskip
Hence we have arrived at one of the CFT manifestations of the bulk
fine-tuning associated to radiative corrections: while the final order
$g^2$ correction to the 4-point function $W^{(2)}$ is small, the
intermediate terms which contribute to the double OPE can be
parametrically larger. In particular, partial sums over some of the
intermediate operators have very large values and are almost cancelled
by large contributions from other operators. All these factors
conspire to give a 4-point function which is only of order $g^2$.
\medskip
In this example the expansion
in \doubleope\ would not seem natural\foot{In the sense that the final
value of the sum over ${\cal A}$ is much smaller than individual
terms.} to a CFT observer.
\medskip

\newsec{The cosmological constant fine-tuning}

After this scalar toy-example let us come back to the fine-tuning of the
cosmological constant. As we explained before, the statement that the bulk
cosmological constant is small (i.e. that the AdS space is large in
units of the Planck scale) means that the correlation functions of
gravitons must be appropriately suppressed. In the CFT this means
that all connected n-point functions of stress-energy tensors have a
specific scaling. In particular
$$
\langle T(x_1)....T(x_n)\rangle_{con} \sim c
$$
for all $n$. Here we have dropped Lorentz indices from the
stress-energy tensors and we have suppressed the $x$-dependence of the
correlator on the RHS. Alternatively we can redefine the stress tensor
as $\widetilde{T}=T/\sqrt{c}$ so that the 2-point function is of order
1. Then the statement of a small cosmological constant can be
understood as the condition
\eqn\ttscale{
\langle \widetilde{T}(x_1)....\widetilde{T}(x_n)\rangle_{con} \sim c^{(2-n)/2}
} So the question is: if we have a non-supersymmetric CFT with the
property \ttscale, whose bulk dual has a cutoff at the Planck scale
(which in CFT language corresponds to the statement that the
generalized free field sector containts few single trace operators up
to conformal dimension $\Delta_{cut}\sim c^{1/(d-1)}$ - see discussion
around equation \condfine), would a CFT observer conclude that the
theory seems fine-tuned?
\medskip
The discussion now continues as before. By considering the 4-point function of
gravitons and the contribution to it by, for example, a fermion loop,
we find that the 1-loop diagram would lead to a violation
of \ttscale. This is avoided by a counterterm diagram, of the form of
a 4-graviton contact interaction.  The two contributions, i.e. the
1-loop diagram and the counterterm, each violate the scaling \ttscale,
but their sum is parametrically smaller and consistent with \ttscale.
\medskip
The boundary manifestation of the fine-tuning is then parallel to what
 was discussed in section 2.2 and in section 4 for a scalar field. The
 correlator
 $\langle \widetilde{T}(x_1) \widetilde{T}(x_2) \widetilde{T}(x_3) \widetilde{T}(x_4)\rangle$
 is of order $1/c$. However in the double OPE we would encounter
 individual conformal blocks, or partial sums over blocks, which are
 parametrically larger and which in the end cancel among
 themselves. These conformal blocks correspond to the exchange of
 double trace operators of the form $:{\cal O}
 (\partial^2)^n \partial_1...\partial_l {\cal O}:$, where ${\cal O}$ can be the operator dual to any of the propagating fields in the bulk (bosonic or fermionic). Each of the corresponding conformal blocks comes with a coefficient of order
 $1/c$ but the sum over $n,l$ leads to a parametrically larger
 value. This is cancelled by conformal blocks corresponding to the
 exchange of operators of the form $:T\partial..\partial T:$, which
 are the boundary dual of the c.c. type counterterm Witten diagram depicted in figure 3.
  Hence in such a CFT the conformal block expansion would seem to be
 fine-tuned.
  
\newsec{Fine-tuning of scalar masses}

In this section we want to discuss another example of fine-tuning, the
one that has to do with the masses of scalar fields. These correspond
to relevant operators and in general there are no symmetries to
protect them from receiving radiative corrections. This is of course
the analogue of the usual hierarchy problem for the Higgs mass in the
Standard Model.

\subsec{\bf The bulk picture}

Let us consider AdS$_4$ and start with the free theory consisting of a
scalar field $\phi$ of mass $m$ and a fermion $\psi$ of mass
$m_f$. The bulk scalar is dual to a scalar single-particle operator
${\cal O}$ of conformal dimension $\Delta$ given by \massdim\ and the
bulk fermion is dual to a fermionic operator $\Psi$. We assume that
the cutoff of the theory is at a scale $M_{cut}$. We turn on a
Yukawa-coupling type of interaction
$$
V= g\, \phi \psi \overline{\psi} + h.c.
$$
At first order in $g$ this turns on a 3-point functions between ${\cal
O}$ and two fermionic operators $\Psi$.

\fig{Yukawa interaction and scalar self-energy.}
{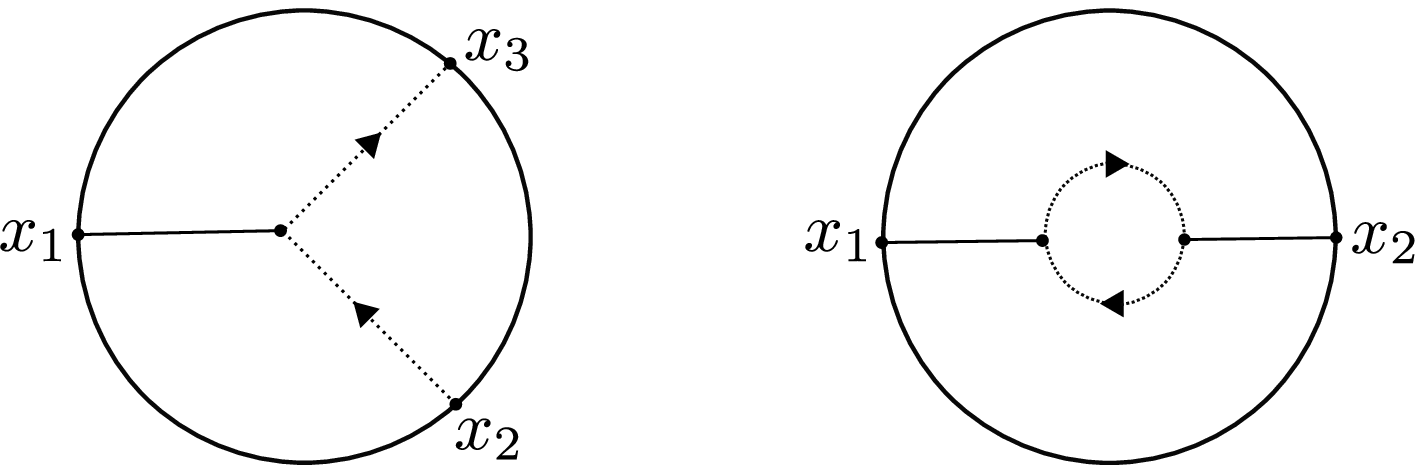}{4.truein}
\figlabel{\fermions}

At second order in $g$ we also have a 1-loop correction to the
self-energy of the scalar $\phi$ shown in figure \fermions. This
diagram will lead to a correction to the tree level mass of the boson
$\phi$.  In order to estimate
the dependence of the diagram on the UV cutoff $M_{cut}$ we can use
the naive flat-space counting. Each of the fermion propagators
contributes a power of $p^{-1}$, where $p$ is the momentum in the
loop, and the integration a power of $p^4$. So we find that this
diagram goes like
$$
\delta m^2\,\approx \,g^2 M_{cut}^2 \,\,+\,\, {\rm subleading}
$$
This is the usual quadratic divergence of the self-energy of scalars
in four dimensions.
\medskip
This diagram implies that the observed mass of the boson would get a
very large contribution of order $g^2 M_{cut}^2$ from the 1-loop
diagram. If however the observed mass is small, then the bulk observer
would attribute it to a cancellation between the 1-loop diagram and a
``mass counterterm''.
\medskip
The fine-tuning is the statement that one would have to balance a tree
level mass with a 1-loop correction to it, both of which are of order
$g^2 M_{cut}^2$, in order to produce a very light mass $m^2$ in the
end.

\subsec{\bf The CFT interpretation}

On the CFT side the meaning of the self energy diagram is as
follows. In the free theory $(g=0)$ the conformal primaries are ${\cal
O}, \Psi$ and their ``mutli-particle'' composites. These operators
have the structure of a freely generated Fock space. When we turn on a
small interaction we have to find the new eigenstates of the
Hamiltonian (on the sphere) i.e. new eigenvalues (conformal
dimensions) and new eigenvectors (conformal primaries) as discussed in
section 3.2.
\medskip
The radiative shift in the mass of the scalar $\phi$ can be understood
as a correction to the energy of the dual state on the sphere, as
computed in perturbation theory in $g$. In this language the
interaction can be written as
\eqn\effint{
V = g V_1 + g^2 V_2 + ...  } where $V_1$ is the interaction from the
tree level Yukawa vertex and $V_2$ contains the counterterms.
\medskip
The conformal dimension of ${\cal O}$ can be written as
$$
\Delta_{\cal O}(g) = \Delta_{\cal O}^{0} + g \delta \Delta_{\cal O}^{(1)} + 
g^2 \delta \Delta_{\cal O}^{(2)} +...
$$
The correction to the dimension of ${\cal O}$ can be computed to first
order in $g$ as

$$
\delta \Delta_{\cal O}^{(1)} =  \langle {\cal O}| V_1 |{\cal O}\rangle
$$
and to order $g^2$ we have
\eqn\secondshift{
\delta \Delta_{\cal O}^{(2)} =   \sum_{\chi,\Delta_{\chi}\neq \Delta_{\cal O}} {\langle {\cal O}|V_1|\chi\rangle \langle \chi|V_1 |{\cal O}\rangle \over \Delta_{\cal O}-\Delta_\chi}+  \langle {\cal O}| V_2 |{\cal O}\rangle
}
The first term comes from the 2nd order perturbation theory expansion
from $V_1$ while the second term from the 1st order perturbation
theory from $V_2$. The sum over $\chi$ is over all other states. In
our particular case the only states with a nontrivial matrix element
are 2-particle states made out of 2 fermions. The sum goes up to the
cutoff $M_{cut}$.
\medskip
The statemet of the fine-tuning is that the first term in \secondshift\ is of order
$M_{cut}^2$, the second term is also of the order $M_{cut}^2$ but
their sum is only of order $m^2 \ll M_{cut}^2$.
\medskip
To summarize: the mass hierarchy problem manifests itself in the CFT
as a puzzle of why there are single-particle conformal primary
operators with small conformal dimension. The naive perturbation
theory of the CFT data predicts a very large correction to the
dimensions which can only be cancelled by a fine-tuning of the
parameters.
\newsec{Naturalness from Holography?}

Let us now speculate on the third question that we asked in the
introduction, namely how the fine-tuning in AdS might be ``resolved''
in a natural way in holographic theories.  The main observation is
that while the description of a large central charge CFT in terms of
an effective theory for its light operators (i.e. the generalized free
fields - the gauge singlets, in the case of large $N$ gauge theories)
is useful for some purposes, like for making the locality of the bulk dual manifest, it is not how the underlying QFT actually computes the
correlation functions.
\medskip
To give an example, consider a large $N$ gauge theory in the 't Hooft
limit. Correlation functions of gauge invariant operators can be
computed in terms of double-line Feynman diagrams. That such
correlation functions are suppressed by the right power of $1/N$ is
obvious in the formalism of the double-line diagrams as 't Hooft
demonstrated. This is the ``fundamental'' way in which correlators are
computed.
\medskip
We could also try to understand these correlators solely in terms of 
gauge invariant objects i.e. in terms of exchange of single trace operators.
For this we have to take the correlators and expand them in
intermediate conformal blocks (for simplicity let us assume that the
theory is conformal). Eventually this leads to an effective description of 
correlators of light operators in terms of gauge singlet collective fields.
This description is best suited to be translated to a bulk theory \ElShowkAG.
The point is that, whether the latter description turns out to look fine-tuned
or not, we do not really need to worry because we understand what is
the underlying mechanism responsible for the smallness of the correlators: it is
the large $N$ combinatorics of the underlying Feynman diagrams. This mechanism is
invisible when we express the correlators in terms of the exchange of
gauge singlets.
\medskip
Similarly, the fine-tuning of light scalar masses appears when the
loop corrections to the dimension of single-trace operators is
calculated by the (effective) interaction Hamiltonian \effint. This
Hamiltonian acts of the Hilbert space corresponding to the Fock space
of the ``generalized free fields''. As we mentioned, while it is
useful to describe the theory in terms of these variables it is not
how the QFT actually determines the dimensions of conformal primaries:
the interaction Hamiltonian $V$ in \effint\ is not the full
Hamiltonian of the underlying QFT. The mechanism responsible for the
smallness of the conformal dimensions of scalars may be easy to
understand in terms of the full Hamiltonian $H_{QFT}$ but invisible in terms
of $V$.
\bigskip

\noindent{\bf The case of large $N$ gauge theories}
\bigskip
If large $N$ gauge theories exhibited fine-tuning in their
conformal block expansion, they would be excellent examples where the
fine-tuning could be resolved by an underlying mechanism (the large
$N$ counting) which is invisible in the language of the gauge
invariant operators. Unfortunately, as mentioned earlier, this is not
the case. In large $N$ gauge theories the conformal block expansion is
natural i.e. no fine-tuning is observed in the $1/N$ expansion in
terms of conformal blocks. This is consistent with the fact that large $N$
gauge theories in the 't Hooft limit are expected to be dual to string
theories, i.e. theories where the bulk cutoff is at the order of the
"string scale" which is parametrically lower than the planck scale in
the large $N$ limit (this is related to the Hagedorn growth in the
spectrum of single-trace operators).
\medskip
Even though large $N$ gauge theories do not exhibit the effect that we
want to see (i.e. fine-tuning of the conformal block expansion), we
can still explore whether they hold any surprises for the IR observer
as far as naturalness is concerned. Let us consider $SU(N)$ Yang-Mills
with $N_f$ fundamental quarks in the 't Hooft large $N$ limit, keeping
$N_f$ fixed. It is believed that the theory confines and below the
strong coupling scale $\Lambda_{QCD}$ the spectrum of particles
consists of gauge singlets i.e. glueballs and mesons. If we were low
energy observers we would see all these particles and we might try to
construct an effective action describing their interactions. From the
low energy point of view we would write the effective action by
introducing a field for each particle and the underlying $SU(N)$ color
structure would of course be invisible.
\medskip
Assuming that the quark masses are small\foot{By this we mean of the
order $\Lambda_{QCD}$. If the quark masses are significantly smaller
then we will have approximate $SU(N_f)\times SU(N_f)$ symmetry, whose
axial part is spontaneously broken by chiral symmetry breaking. The
resulting (almost) massless mesons will dominate the IR effective
Lagrangian.}, the masses of the glueballs and mesons will be of order
$\Lambda_{QCD}$. However we know that couplings between glueballs are
suppressed by powers of $1/N$ while those between mesons by powers of
$1/\sqrt{N}$.  In the large $N$ limit there is a hierarchy between
these couplings, which would probably surprise the low energy observer
given that he has no other way of qualitatively distinguishing
glueballs from mesons. Moreover, even among meson interactions it
turns out that various processes are suppressed by factors of $1/N$
depending on how many quark lines have to be drawn in the double-line
diagrams. This is the so-called ``OZI rule"\foot{Okubo-Zweig-Iizuka rule.}. It implies that in the
IR effective action for the mesons certain couplings will be
suppressed by additional powers of $1/N$. This rule can be easily
understood in terms of double-line diagrams (see for
example \WittenKH), but would be mysterious for the IR observer who
works directly with the mesons.
\medskip
The reason that this toy model is not fully satisfactory is that the
cutoff of the effective field theory is $\Lambda_{QCD}$ and all
interactions between gauge singlets are suppressed by powers of
$1/N$. Hence the effects of loops (in the IR effective Lagrangian, not the UV nonabelian theory) are
quite small and cannot destabilize the tree level values. So while the
IR observer would indeed notice large hierarchies between coupling
constants and decay rates, he would not have to worry about
fine-tuning between loop diagrams and counterterms.
\bigskip

\newsec{Discussions}

We discussed the conditions under which a bosonic CFT with large
central charge has a holographic dual with a sharp ``cosmological
constant problem''. We argued that in such a CFT the bulk fine-tuning
manifests itself in terms of an apparent fine-tuning of the $1/N$
expansion of correlators in conformal blocks. Finally we proposed the
idea that while this fine-tuning may be visible when the correlators
are expressed in terms of the exchange of conformal primaries, it may
disappear if the correlators are written in terms of the fundamental
fields of the underlying QFT.
\medskip
If this last possibility is true and if we extrapolate it to our
world, it would imply that the mechanism responsible for the
fine-tuning may be invisible in terms of the
low energy fields (the graviton and the fields of the Standard Model)
but may be simple to understand in terms of the fundamental fields
(i.e. in the ``holographic dual of the universe'').
\medskip
How can we check if this possibility has any chance of being true, at
least in the context of AdS/CFT? The most optimistic scenario would be
to find a specific CFT whose AdS dual has a sharp c.c. problem and to
check how exactly the fine-tuning is resolved on the boundary. Of
course it would be very nice to have such an example for many other
reasons, as it would be a non-perturbatively well defined theory of
AdS quantum-gravity without supersymmetry\foot{As we explained before,
in all known bosonic examples of AdS/CFT the bulk theory has a very
low cutoff and thus does not have a sharp c.c. problem.}.
\medskip
A somewhat easier goal is to find a toy-model which would illustrate
the general logic in a simpler setting.  For instance, it would be nice
to find an example of a QFT where a prediction of effective field
theory (related to the smallness of relevant operators) is invalidated
due to some microscopic mechanism which is invisible in terms of the
effective degrees of freedom in the IR. Or, more likely, a quantum
system (not necessarily a full-fledged QFT) where certain computations
in terms of effective/collective degrees of freedom
of the system give misleading answers relative to the exact
computation in terms of the fundamental degrees of freedom. 
We hope to report on this in the future.
\medskip
The idea that the cosmological constant fine-tuning may be an artifact of
using emergent rather than fundamental variables has also been
discussed by G.E. Volovik \VolovikBH\ in a rather different
context. We would like to thank E. Verlinde for bringing these
papers to our attention and for many other discussions regarding his recent work
\VerlindeHP.

\medskip
Let us finish with some general observations on a related topic, which was also emphasized
in \FitzpatrickZM.
\bigskip
\noindent {\it The meaning of effective field theory and RG-flow in the bulk}
\medskip
After all, in order to find a satisfactory explanation of the
c.c. fine-tuning we have to understand why the predictions of
effective field theory are not completely reliable in some
situations. The idea of naturalness is based on the paradigm of
Wilsonian RG-flow, i.e. that the low-energy effective theory can be
derived by doing RG-flow from a UV theory. From this point of view it
is indeed difficult to explain small coefficients of relevant
operators. However the Wilsonian intuition seems to be in some tension
with the idea of holography where the bulk is an emergent structure:
the low-energy effective field theory in the bulk is not necessarily
determined by RG-flow from a UV theory in the bulk, but
holographically from the boundary QFT, though there may also be some
partial consistency condition with the bulk RG-flow. In order to understand this
better it would be interesting to develop further the boundary
interpretation of the bulk RG-flow.  \fig{It is known that the usual
RG-flow in the boundary QFT can be related to the radial evolution in
the bulk. What is the boundary meaning of the bulk RG-flow?}
{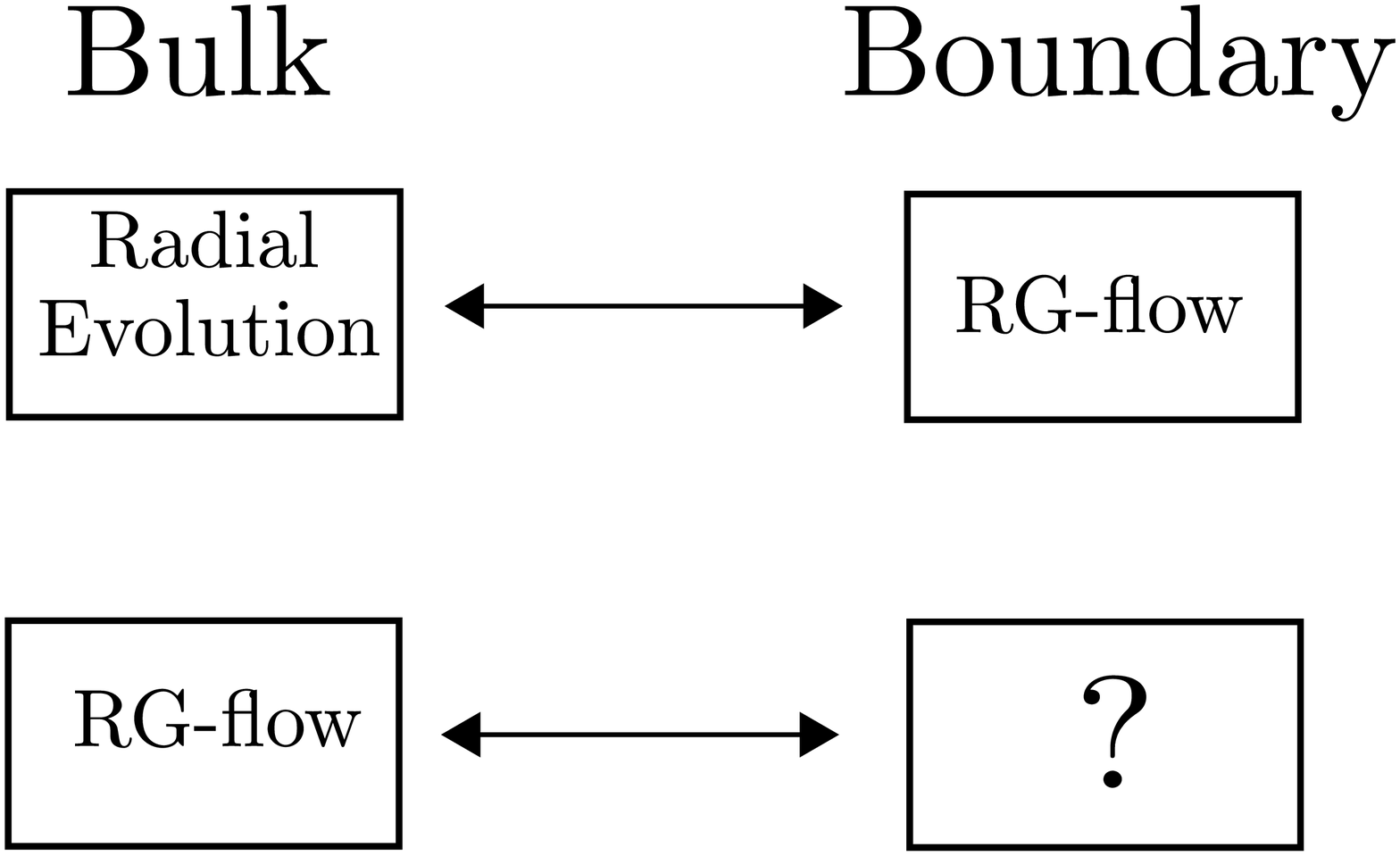}{3.truein}
\figlabel{\disconnected}
\medskip
In the AdS/CFT literature there have been extensive discussions about
the relation of the RG-flow in the boundary field theory to the radial
evolution in the bulk.  This is not what we are talking about. Here we
are referring to RG-flow in the bulk and its meaning in the dual
CFT. Even if the boundary theory is an exact conformal field theory
(no mass scale), the theory in the bulk obviously has certain length
scales, such as the Planck mass and the masses of other particles in
AdS. The masses of these fields are dual to the conformal dimensions
of single-particle operators in the dual CFT. In the bulk it is
perfectly normal to consider the RG-flow in the effective field
theory. This translates into some kind of flow in the direction of
conformal dimensions (see also \FitzpatrickZM). This flow is distinct
from the standard RG-flow in the field theory, which is trivial for
CFTs. Presumably this new type of flow only makes sense in CFTs with a
holographic dual, or in other words, in the sector of the CFT
described by the generalized free fields. It might be interesting to
develop this further, perhaps assuming (for simplicity) that there is
a regime where non-gravitational interactions in the bulk become
parametrically more important than gravity.
\bigskip
\bigskip
\bigskip
\bigskip
\bigskip  
\centerline{\bf Acknowledgments}
\medskip
We would like to  J. de Boer, S. El-Showk, R. Emparan, B. Fiol,
N. Iizuka, G. Pastras, J. Penedones, S. Raju, R. Rattazzi, S. Rychkov,
M. Shigemori and especially S. Minwalla and E. Verlinde for
discussions. We would like to thank the Perimeter Institute for
Theoretical Physics for hospitality during the ``Back to the
Bootstrap'' Workshop. We also thank the theory group in Santiago de
Compostela and the group in Barcelona for hospitality during the
completion of this work.
\listrefs
\end